%% file: CompoNet.tex
\pdfoutput=1
\documentclass[10pt,twocolumn,letterpaper]{article}

\usepackage{iccv}
\usepackage{times}
\usepackage{epsfig}
\usepackage{graphicx}
\usepackage{amsmath}
\usepackage{amssymb}
\usepackage{multirow}
\usepackage{hhline}

\usepackage{color}
\usepackage{caption}
\usepackage{subcaption}

\definecolor{dark_green}{rgb}{0, 0.5, 0}
\newcommand{\rulesep}{\unskip\ \vrule\ }

\newcommand{\OurNet}{CompoNet}
\usepackage[pagebackref=true,breaklinks=true,letterpaper=true,colorlinks,bookmarks=false]{hyperref}

\iccvfinalcopy %

\def\iccvPaperID{1643} %

\ificcvfinal\fi
\begin{document}

\title{CompoNet: Learning to Generate the Unseen by Part Synthesis and Composition}

\author{
Nadav Schor$^1$ \quad Oren Katzir$^1$ \quad Hao Zhang$^2$  \quad Daniel Cohen-Or$^1$ \\
Tel Aviv University$^1$ \quad Simon Fraser University$^2$\\
{\tt\small nadav.schor@cs.tau.ac.il} \quad {\tt\small orenkatzir@mail.tau.ac.il} \quad {\tt\small haoz@cs.sfu.ca} \quad {\tt\small dcor@tau.ac.il}\\
}

\twocolumn[{
\maketitle

}]

``For object recognition, the visual system decomposes shapes into parts, $\ldots$, parts with their descriptions and spatial relations provide a first index into a memory of shapes.''
\vspace{-15pt}
\begin{flushright}
--- Hoffman \& Richards~\cite{hoffman1984}
\end{flushright}

\vspace{10pt}

\begin{abstract}
	\input{abstract}
\end{abstract}

\input{intro}

\input{related}

\input{algorithm}
\input{implementation}
\input{results}

\input{future}

{\small
\bibliographystyle{ieee}
\bibliography{CompoNet}
}

\clearpage
\appendix
\input{supp}

\end{document}

%% file: abstract.tex
Data-driven generative modeling has made remarkable progress by leveraging the power of deep neural networks. A reoccurring challenge is how to enable a model to generate a rich variety of samples from the {\em entire\/} target distribution, rather than only from a distribution confined to the training data. In other words, we would like the generative model to go beyond the observed samples and learn to generate ``{\em unseen\/}'', yet still plausible, data. In our work, we present \OurNet{}, a generative neural network for 2D or 3D shapes that is based on a part-based prior, where the key idea is for the network to synthesize shapes by varying both the shape parts and their {\em compositions\/}. Treating a shape not as an unstructured whole, but as a (re-)composable set of deformable parts, adds a combinatorial dimension to the generative process to enrich the diversity of the output, encouraging the generator to venture more into the ``unseen". We show that our part-based model generates richer variety of plausible shapes compared with baseline generative models. To this end, we introduce two quantitative metrics to evaluate the diversity of a generative model and assess how well the generated data covers both the training data and unseen data from the same target distribution. Code is available at \url{https://github.com/nschor/CompoNet}.

%% file: intro.tex
\section{Introduction}
\label{sec:intro}

\begin{figure}[t]
	\centering
	\includegraphics[width=\linewidth]{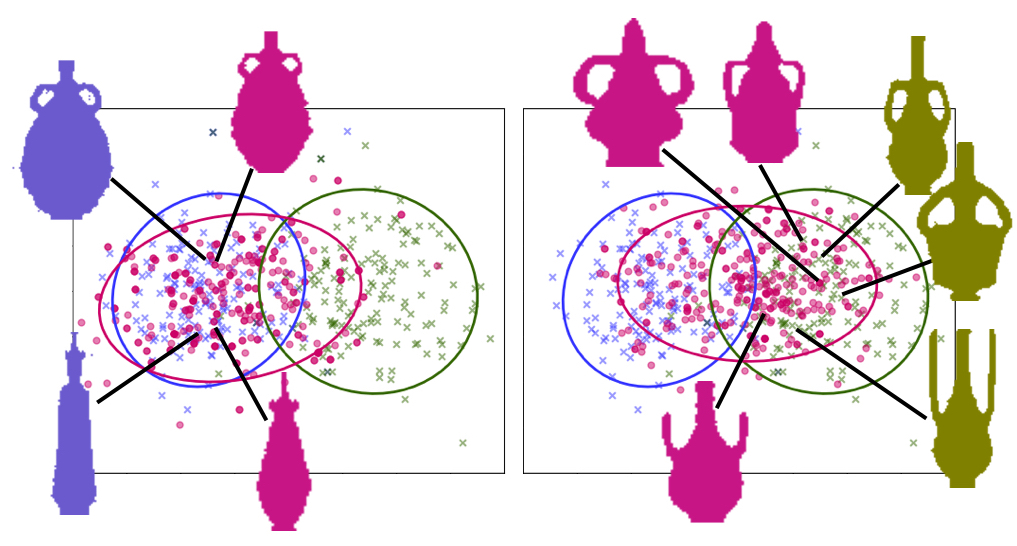}
   	\begin{minipage}[c]{0.48\linewidth} \hspace{0.25cm} \centering (a) Baseline  \end{minipage}
 \begin{minipage}[c]{0.48\linewidth} \centering \hspace{-0.35cm} (b) Part-based generation \end{minipage}
	\caption{\OurNet{}, our part-based generative model (b) covers the ``unseen'' data significantly more than the baseline (a). Generated data (pink dots) by two methods are displayed over training data (purple crosses) and unseen data (green crosses) from the same target distribution. The data is displayed via PCA over a classifier feature space, with the three distributions summarized by ellipses, for illustration only. A few samples of training, unseen, and generated data are displayed to reveal their similarity/dissimilarity.}
	\label{fig:teaser}
\end{figure}

Learning generative models of shapes and images has been a long standing research problem in visual computing. Despite the remarkable progress made, an inherent and reoccurring limitation still remains: a generative model is often only as good as the given training data, as it is always trapped or bounded by the empirical distribution of the observed data. More often than not, what can be observed is not sufficiently expressive of the true target distribution. Hence, the generative power of a learned model should not only be judged by the plausibility of the generated data as confined by the training set, but also its {\em diversity}, in particular, by the model's ability to generate plausible data that is sufficiently far from the training set. Since the target distribution which encompasses both the observed and unseen data is unknown, the main challenge is how to effectively train a network to learn to generate the ``unseen'', without making any compromising assumption about the target distribution. Due to the same reason, even evaluating the generative power of such a network is a non-trivial task.

\begin{figure*}[t]
	\centering
	\includegraphics[width=.96\linewidth]{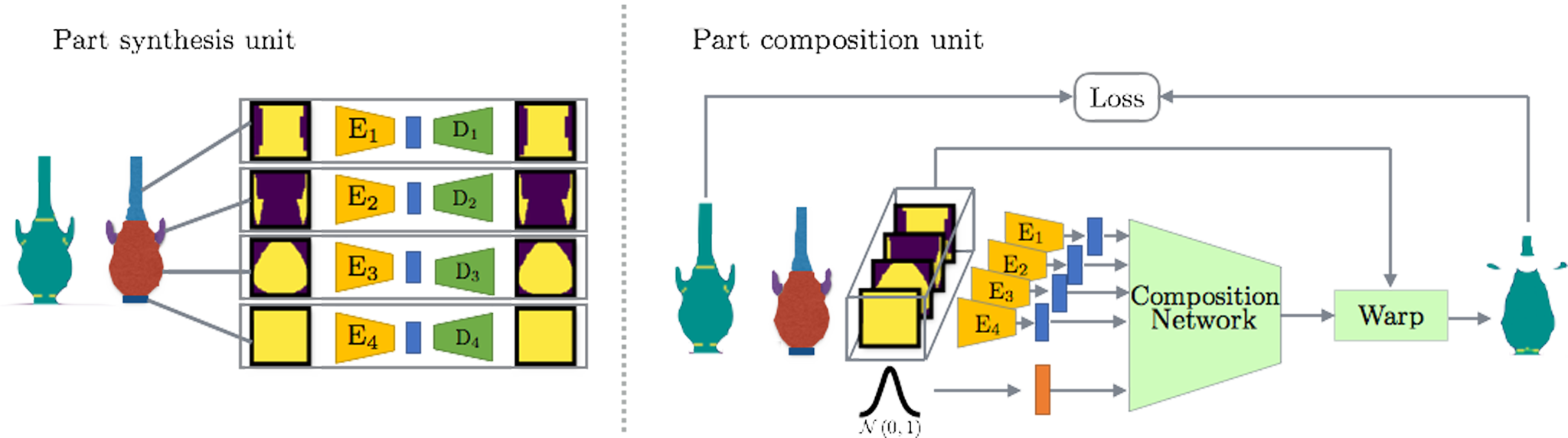}
	\caption{The two training units of \OurNet{}, our part-based generative model: 
The first unit, the {\em part synthesis unit\/}, consists of parallel generative AEs; an independent AE for each semantic part of the shape. The second unit, the {\em part composition unit\/}, learns to compose the encoded parts. We use the pre-trained part encoders from the part synthesis unit. Then, a noise vector $z$, is concatenated to the parts' latent representation and fed to the composition network, which outputs transformation parameters per part. The parts are then warped and combined to generate the entire input sample.}
	\label{fig:train_arch}
\end{figure*}

We believe that the key to generative diversity is to enable more drastic changes, i.e., non-local and/or structural transformations, to the training data. At the same time, such changes must be within the confines of the target data distribution. In our work, we focus on generative modeling of 2D or 3D {\em shapes\/}, where the typical modeling constraint is to produce shapes belonging to the same category as the exemplars, e.g., chairs or vases. We develop a generative deep neural network based on a \emph{part-based\/} prior. That is, we assume that shapes in the target distribution are composed of parts, e.g., chair backs or airplane wings. %
The network, coined \OurNet{}, is designed to synthesize novel parts, {\em independently\/}, and compose them to form a complete %
shape.

It is well-known that object recognition is intricately tied to reasoning about parts and part relations~\cite{hoffman1984,thompson1992}. Hence, building a generative model based on varying parts
and their compositions, while respecting category-specific part priors, is a natural choice and also facilitates grounding of the generated data to the target object category. More importantly, treating a shape as a (re-)composable set of parts, instead of a whole entity, adds a {\em combinatorial\/} dimension to the generative model and improves its diversity. By synthesizing parts independently and then composing them, our network enables both part variation and novel combination of parts, which induces non-local and more drastic shape transformations. Rather than sampling only a single distribution to generate a whole shape, our generative model samples both the geometric distributions of individual parts and the combinatorial varieties arising from part compositions, which encourages the generative process to venture more into the ``unseen'', as shown in Figure~\ref{fig:teaser}.

While the part-based approach is generic and not strictly confined to specific generative network architecture, we develop a {\em generative autoencoder\/} %
to demonstrate its potential. Our generative AE consists of two parts. In the first, we learn a distinct part-level generative model. In the second stage, we concatenate these learned latent representation with a random vector, to generate a new latent representation for the entire shape. These latent representations are fed into a conditional parts compositional network, which is based on a spatial transformer network (STN) \cite{jaderberg2015spatial}.%

We are not the first to develop deep neural networks for part-based modeling. Some networks learn to compose images \cite{lin2018stgan,azadi2018compositional} or 3D shapes \cite{kalogerakis2012probabilistic,Chaudhuri2011,zhu2018scores}, by combining {\em existing\/} parts sampled from a training set or provided as input to the networks. In contrast, our network is fully generative as it learns both novel part synthesis and composition. Wang \etal~\cite{wang2018g2lgan} train a generative adversarial network (GAN) to produce semantically segmented 3D shapes and then refine the part geometries using an autoencoder network. Li \etal~\cite{li2017grass} train a VAE-GAN to generate structural hierarchies formed by bounding boxes of object parts and then fill in the part geometries using a separate neural network. Both works take a coarse-to-fine approach and generate a rough 3D shape holistically from a noise vector. In contrast, our network is trained to perform both part synthesis and part composition (with noise augmentation); see Figure~\ref{fig:train_arch}. Our method also allows the generation of more {\em diverse parts\/}, since we place less constraints per part while holistic models are constrained to generate all parts at once.

We show that the part-based \OurNet{} produces plausible outputs that better cover unobserved regions of the target distribution, compared to baseline approaches, e.g.,~\cite{achlioptas2018learning}. This is validated over random splits of a set of shapes belonging to the same category into a ``seen'' subset, for training, and an ``unseen'' subset.
In addition, to evaluate the generative power of our network relative to baseline approaches, we introduce two quantitative metrics to assess how well the generated data covers both the training data and the unseen data from the same target distribution.

\begin{figure}[t]
	\centering
	\includegraphics[width=\linewidth]{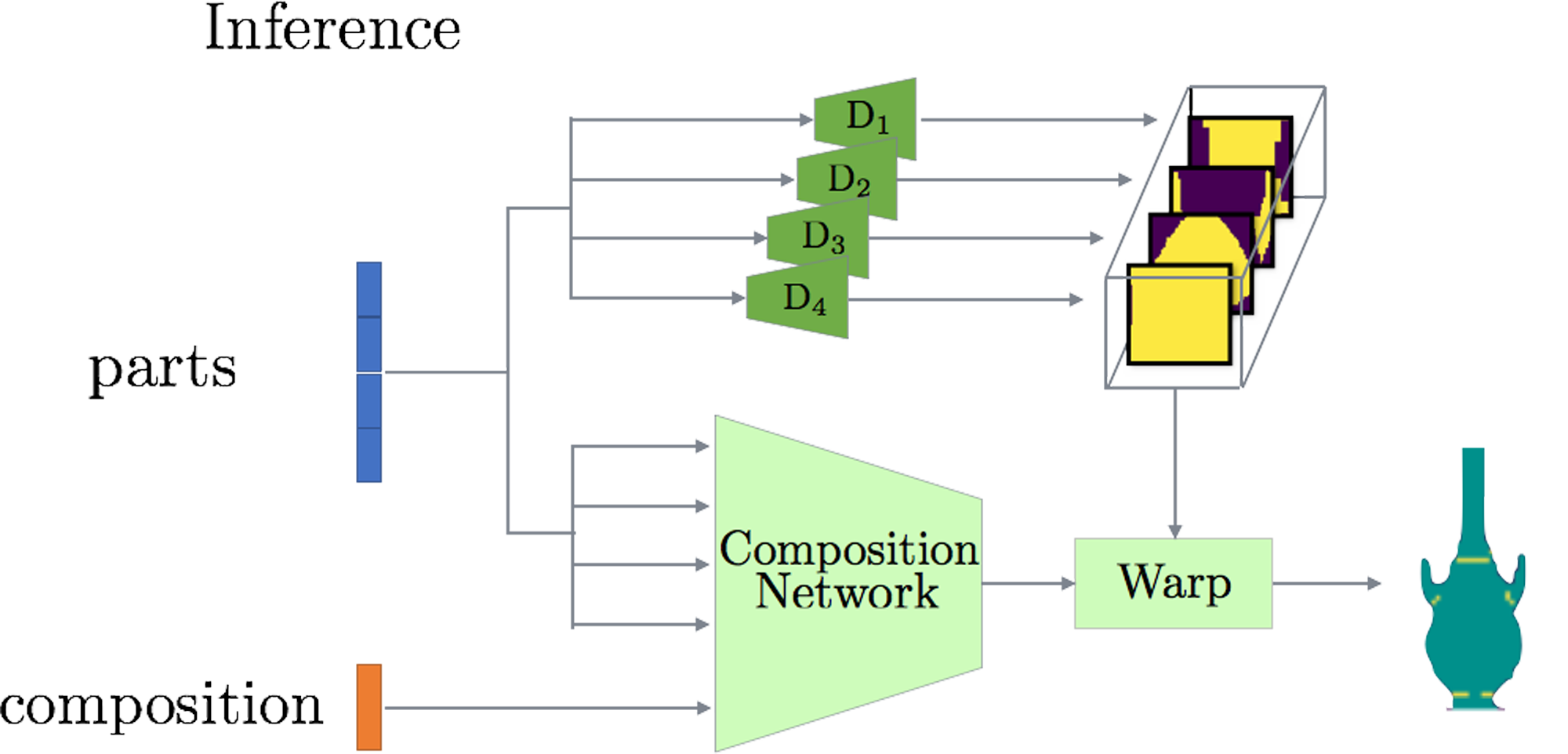}
	\caption{Novel shape generation at inference time. We randomly sample the latent spaces for shape parts and part composition. Using the pre-trained part-decoders and the composition network, we generate novel parts and then warp them to produce a coherent whole shape. 
}
	\label{fig:CompNet}
\end{figure}

%% file: related.tex
\section{Background and Related Work}
\label{sec:related}

\paragraph{Generative neural networks.}
In recent years, generative modeling has gained much attention within the deep learning framework. Two of the most commonly used deep generative models are variational auto-encoders (VAE)~\cite{kingma2014auto} and generative adversarial networks (GAN)~\cite{goodfellow2014generative}.
Both methods have made remarkable progress in image and shape generation problems~\cite{wu2016learning, isola2017image,radford2015unsupervised, zhu2016generative, wang2016generative, yan2016attribute2image, wang2018g2lgan}.

Many works are devoted to improve the basic models and their training. In~\cite{gulrajani2017improved, mao2017least, berthelot2017began}, new cost functions are suggested to achieve smooth and non-vanishing gradients.
Sohn \etal\cite{sohn2015learning} and Odena \etal\cite{odena2016semi} proposed conditional generative models, based on VAE and GAN, respectively. Hoang \etal~\cite{hoang2017multi} train multiple generators to explore different modes of the data distribution. Similarly, MIX$+$GAN~\cite{arora2017generalization} uses a mixture of generators to improve diversity of the generated distribution, while a combination of multiple discriminators and a single generator aims at constructing a stronger discriminator to guide the generator. GMAN~\cite{durugkar2016generative} explores an array of discriminators to boost generator learning. Some methods \cite{iizuka2017globally,liu2018image,yu2018generative} use a global discriminator together with multiple local discriminators. 

Following PointNet ~\cite{qi2017pointnet}, a generative model that works directly on point clouds was proposed. Achlioptas \etal~\cite{achlioptas2018learning} proposed an AE+GMM generative model for point clouds, which is considered state-of-the-art.

Our work is orthogonal to these methods. We address the case where the generator is unable to generate other valid samples since they are not well represented in the training data. We show that our part-based priors can assist the generation process and extend the generator's capabilities.

\vspace{-8pt}

\paragraph{Learning-based shape synthesis.}
Li \etal~\cite{li2017grass} present a top-down and structure-oriented approach for 3D shape generation. They learn symmetry hierarchies~\cite{wang2011} of shapes with an autoencoder and generate variations of these hierarchies using an VAE-GAN. The nodes of the hierarchies are independently instantiated with parts. However, these parts are not necessarily connected and their aggregation does not form a coherent connected shape. In our work, the shapes are generated coherently as a whole, and special care is given to the inter-parts relation and their connectivity.

Most relevant to our work is the shape variational auto-encoder by Nash and Williams~\cite{nash2017shape}, where a point-cloud based autoencoder is developed to learn a low-dimensional latent space. Then novel shapes can be generated by sampling vectors in the learned space. Like our method, the generated shapes are segmented into semantic parts. In contrast however, they require a one-to-one dense correspondence among the training shapes, since they represent the shapes as an order vector. Their autoencoder learns the overall (global) 3D shapes with no attention to the local details. Our approach pays particular attention to both the generated parts and their composition.

\vspace{-8pt}

\paragraph{Inverse procedural modeling}
aims to learn a generative procedure from a given set of exemplars. Some recent works, e.g.,~\cite{ritchie2016, yumer2015, sharma2018}, have focused on developing neural models, such as autoencoders, to generate the shape synthesis procedures or programs. However, current inverse procedural modeling methods are not designed to generate unseen data that are away from the exemplars.

\vspace{-8pt}

\paragraph{Assembly-based synthesis.}
The early and seminal work of Funkhouser \etal~\cite{funkhouser2004modeling} composes new shapes by retrieving relevant shapes from a repository, extracting shape parts, and gluing them together. Many follow-up works~\cite{bokeloh2010, talton2012learning, chaudhuri2010data, kalogerakis2012probabilistic,xu2012fit, kim2013learning, fish2014meta,huang2015analysis} improve the modeling process with more sophisticated techniques that consider the part relations or shape structures, e.g., employing Bayesian networks or modular templates. We refer to recent surveys~\cite{mitra2013star,mitra2013sig} for an overview of these and related works.

In the image domain, recent works~\cite{lin2018stgan, azadi2018compositional} develop neural networks to assemble images or scenes from existing components. These works utilized an STN \cite{jaderberg2015spatial} to compose the components to a coherent image/scene. In our work, an STN is integrated as an example for prior information regarding the data generation process. In contrast to previous works, we first synthesize parts using multiple generative AEs and then employ an STN to compose the parts. 

Recent concurrent efforts~\cite{dubrovina2019composite,li2019learning} also propose deep neural networks for shape modeling using a part-based prior, but on voxelized representations. Dubrovina \etal~\cite{dubrovina2019composite} encode shapes into a factorized embedding space, where shape composition and decomposition become simple linear operations on the embedding coordinates, allowing both shape reconstruction and part exchange. While this work was not going after generative diversity, the network of Li \etal~\cite{li2019learning} also combines part generation with assembly. Their results reinforce our premise that shape generation using part synthesis and composition does improve diversity, which is 
measured using inception scores in their work.

%% file: algorithm.tex
\section{Method}
\label{sec:method}

In this section, we present \OurNet{}, our generative model which learns to synthesize shapes that can be represented as a composition of distinct parts. At training time, every shape is pre-segmented to its semantic parts, and we assume that the parts are independent of each other. Thus, every combination of parts is valid, even if the training set may not include it. As shown in Figure~\ref{fig:train_arch}, \OurNet{} consists of two units: a generative model of parts and a unit that combines the generated parts into a global shape.

\subsection{Part synthesis unit}

We first train a generative model that estimates the marginal distribution of each part separately. In the 2D case, we use a standard VAE as the part generative model, and train an individual VAE for each semantic part. Thus, each part is fed into a different VAE and is mapped onto a separate latent distribution. The encoder consists of several convolutional layers followed by Leaky-ReLU activation functions. The final layer of the encoder is a fully connected layer producing the latent distribution parameters. Using the reparameterization trick, the latent distribution is sampled and decoded to reconstruct each individual input part. The decoder mirrors the encoder network, applying a fully connected layer followed by transposed convolution layers with ReLU non-linearity functions. 
In the 3D case, we borrow an idea from Achlioptas \etal~\cite{achlioptas2018learning}, and replace the VAE with an AE+GMM, where we approximate the latent space of the AE by using a GMM. The encoder is based on PointNet~\cite{qi2017pointnet} architecture and the decoder consists of fully-connected layers. The part synthesis process is visualized in Figure~\ref{fig:train_arch}, part synthesis unit.

Once the part synthesis unit is trained, the part encoders are fixed, and are used to train the part composition unit.

\subsection{Parts composition unit}

This unit composes the different parts into a coherent shape. Given a shape and its parts, where missing parts are represented by null shapes (i.e., zeros), the pre-trained encoders encode the corresponding parts (marked in blue in Figure~\ref{fig:train_arch}).
At training time, these generated codes are fed into a composition network which learns to produce transformation parameters per part (scale and translation), such that the composition of all the parts forms a coherent complete shape. The loss measures the similarity between the input shape and the composed shape. We use Intersection-over-Union (IoU) as our metric in the 2D domain, and Chamfer distance for the 3D domain, where the Chamfer distance is given by
\begin{eqnarray}
d_C(Q,P) = \sum_{q\in Q} \min_{p \in P} (q-p)^2 + \sum_{p\in P} \min_{q \in Q} (p-q)^2,
\end{eqnarray}
where $P$ and $Q$ are point clouds which represent the 3D shapes.
Note that the composition network yields a set of affine (similarity) transformations, which are applied on the input parts, and does not directly synthesizes the output. 

The composition network does not learn the composition based solely on part codes, but also relies on an input noise vector. This network is another generative model on its own, generating the scale and translation from the noise, conditioned on the codes of the semantic parts. This additional generative model enriches the variation of the generated shapes, beyond the generation of the parts.

\subsection{Novel shape generation}

At inference time, we sample the composition vector from a normal distribution. In the 2D case, since we use VAEs, we sample the part codes from normal distribution as well. For 3D, we sample the code of each part from its GMM distribution, %
randomly sampling one of the Gaussians. When generating a new shape with a missing part, we use the embedding of the part's null vector, and synthesize the shape from that compound latent vector; see Figure \ref{fig:CompNet}. We feed each section of the latent vector representing a part to its associated pre-trained decoder from the part synthesis unit, to generate novel parts. In parallel, the entire shape representation vector is fed to the composition network to generate scale and translation parameters for each part. The synthesized parts are then warped according to the generated transformations and combined to form a novel shape. 

%% file: implementation.tex
\section{Architecture and implementation details}
\label{sec:implementation}

The backbone architecture of our part based synthesis is an AE: VAE for 2D and AE+GMM for 3D.

\subsection{Part-based generation}

\paragraph{2D Shapes.}
The input parts are assumed to have a size of $64 \times 64 \times 1$. We denote $C(k)$ ($TC(k)$) as a 2D convolution (transpose convolution) layer with $k$ filters of size $5 \times 5$ and stride $2$, followed by batch normalization and a leaky-ReLU (ReLU) activation. A fully-connected layer with $k$ outputs is denoted by $L(k)$. The encoder takes a 2D part as input and has the structure of $C(8)-C(16)-C(32)-C(64)-L(10)$. 
The decoder mirrors the encoder as $L(1024)-TC(32)-TC(16)-TC(8)-TCS(1)$, where in the last layer, $TCS$, we omitted batch normalization and replaced ReLU by a Sigmoid activation. The output of the decoder is equal in size to the 2D part input, ($64 \times 64 \times 1$).
We use an Adam optimizer with learning rate $= 2e^{-4}$, $\beta_1=0.5$ and $\beta_2=0.999$. The batch size is set to $64$.

\vspace{-8pt}

\paragraph{3D point clouds.}
Our input parts are assumed to have a fixed number of point for each part. Different parts can vary in number of points, but this becomes immutable once training has started. We used $400$ points per part. We denote $MP$ as a feature-wise max-pooling layer and $1DC(k)$ as a 1D convolution layer with $k$ filters of size $1$ and stride $1$, followed by a batch normalization layer and a ReLU activation function. The encoder takes a part with $400 \times 3$ as input. The encoder structure is $1DC(64)-1DC(64)-1DC(64)-1DC(128)-1DC(64)-MP$.
The decoder consist of fully-connected layers. We denote $L(k)$ to be a fully-connected layer with $k$ output nodes, followed by a batch normalization layer and a ReLU activation function. The decoder takes a latent vector of size $64$ as input. The decoder structure is $L(256)-L(256)-LC(400 \times 3)$, where in the last layer, $LC$, we omitted the batch-normalization layer and the ReLU activation function. The output of the decode is equal in size to the input ($400 \times 3$).
For each AE, we use a GMM with $20$ Gaussians, to model their latent space distribution.
We use an Adam optimizer with learning rate $= 0.001$, $\beta_1=0.9$ and $\beta_2=0.999$. The batch size is set to $64$.

\subsection{Part composition}

\paragraph{2D.} The composition network encodes each semantic part by the associated pre-trained VAE encoder, producing a $10$-dim vector for each part. The composition noise vector is set to be $8$-dim. The part codes are concatenated together with the noise, yielding a $48$-dim vector. The composition network structure is $L(128)-L(128)-L(16)$. Each fully connected layer is followed by a batch normalization layer, a ReLU activation function, and a Dropout layer with keep rate of $0.8$, except for the last layer. The last layer outputs a $16$-dim vector, four values per part. These four values represent the scale and translation in the $x$ and $y$ axes.
We use the grid generator and sampler, suggested by \cite{jaderberg2015spatial}, to perform differential transformation. The scale is initialized to $1$ and the translation to $0$. We use a per-part IoU loss and an Adam optimizer with learning rate $= 0.001$, $\beta_1=0.9$ and $\beta_2=0.999$. The batch size is set $64$.

\vspace{-8pt}

\paragraph{3D.}
The composition network encodes each semantic part by the associated pre-trained AE encoder, producing a $64$-dim vector for each part. The composition noise vector is set to size $16$. The parts codes are concatenated together with the noise vector, yielding a $272$-dim vector. The composition network structure is $L(256)-L(128)-L(24)$. Each fully connected layer is followed by a batch normalization layer and a ReLU activation function, except for the last layer. The last layer outputs a $24$-dim vector, six values per part. These six values represent the scale and translation in the $x$, $y$ and $z$ axes. The scale is initialized to $1$ and the translation to $0$. We then reshape the output vector to match an affine transformation matrix:
\begin{eqnarray}
   M=
  \left[ {\begin{array}{cccc}
   s_x & 0 & 0 & t_x \\
   0 & s_y & 0 & t_y \\
   0 & 0 & s_z & t_z \\
  \end{array} } \right]
\end{eqnarray}
The task of performing an affine transformation on point clouds is easy, we simply concatenate $1$ to each point $(x,y,z,1)$ and multiply the transformation matrix with each point.
We use Chamfer distance loss and an Adam optimizer with learning rate $= 0.001$, $\beta_1=0.9$ and $\beta_2=0.999$. The batch size is set $64$.

%% file: results.tex
\section{Results and evaluation}
\label{sec:results}

In this section, we analyze the results of applying our generative approach to 2D and 3D shape collections.

\begin{figure*}[t]
	\centering
	\includegraphics[width=.75\linewidth]{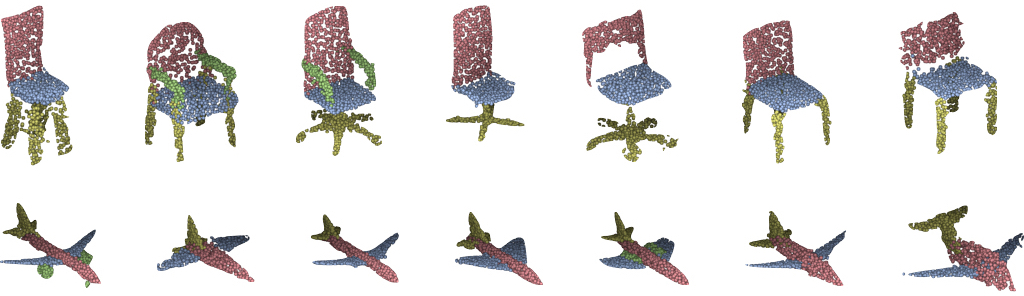}
	\caption{Representative samples of 3D shapes generated by our part-based generative network, \OurNet{}.}
	\label{fig:gallery_results}
\end{figure*}

\subsection{Datasets}

\paragraph{Projected COSEG.}
We used the COSEG dataset \cite{sidi2011coseg} which consists of $300$ vases, segmented to four different semantic labels: top, handle, body and base (each vase may or may not contain any of these parts). Similar to the projection procedure in \cite{fish2016structure}, each vase is projected from the main view to constitute a collection of 300 silhouettes of size $64 \times 64$, where each semantic part is stored in a different channel. In addition, we create four sets, one per part. The parts are normalized by finding their axis-aligned bounding box and stretching it to a $64 \times 64$ resolution. 
\vspace{-7pt}

\paragraph{Shape-Net.}
For 3D data, we chose to demonstrate our method on point clouds taken from ShapeNet part dataset \cite{yi2016scalable}. We chose to focus on two categories: chairs and airplanes. Point clouds, compared to 3D voxels, enable higher resolution while keeping the model complexity relatively low.  Similar to the 2D case, each shape is divided into its semantic parts (Chair: legs, back, seat and arm-rests, Airplane: tail, body, engine and wings). We first normalize each shape to the unit square. We require an equal number of points $N$ in each point cloud, thus, we randomly sample each part to $N=400$ points. If a part consists of $M<N$ points, we randomly duplicate $N-M$ of its points (since our non-local operation preforms only max global pooling, the duplication of points has no affect on the embedding of the shape). This random sampling process occurs every epoch. For consistency between the shape and its parts, we first normalize the original parts to the unit square, and only then sample (or duplicate) the same points that were selected to generate the complete sampled shape.
\vspace{-7pt}

\paragraph{\emph{Seen} and \emph{Unseen} splits.}
To properly evaluate the diversity of our generative model, we divide the resulting collections into two subsets: (i) training (\emph{seen}) set and (ii) \emph{unseen} set. The term \emph{unseen} emphasizes that unlike the nominal division into train and test sets, the unseen set is not represented well in the training set. Thus, there exists an unbridgeable gap, by an holistic approach, between the unseen set and the training set. To avoid bias during evaluation, we preform several random splits for each seen-unseen split percentage (e.g., $15\%$-$85\%$ seen-unseen in the 3D case; see Table~\ref{table:classifier_res}). In the 2D case, since the dataset is much smaller, we used $50\%$-$50\%$ split between the training and unseen sets. In both cases, the unseen set is used to evaluate the ability of a model to generate diverse shapes.

\subsection{Baselines}
\label{sec: Baselines}

For 2D shapes, we use a naive model - a one-channel VAE. Its structure is identical to the part VAE with a latent space of $48$-dim. We feed it with a binary representation of the data (a silhouette) as input. We use an Adam optimizer with learning rate $= 2e^{-4}$, $\beta_1=0.5$ and $\beta_2=0.999$. The batch size is set to $64$.
In the 3D case, we use two baselines; (i) a WGAN-GP~\cite{gulrajani2017improved} for point clouds and (ii) an AE+GMM model~\cite{achlioptas2018learning}, which generates remarkable 3D point cloud results. %
We train the baseline models using our $400$-points per part data set ($1600 \times 3$ per shape). We use \cite{achlioptas2018learning} official implementation and parameters, which also includes the WGAN-GP implementation. 

\begin{figure}
	\centering
	\includegraphics[width=\linewidth]{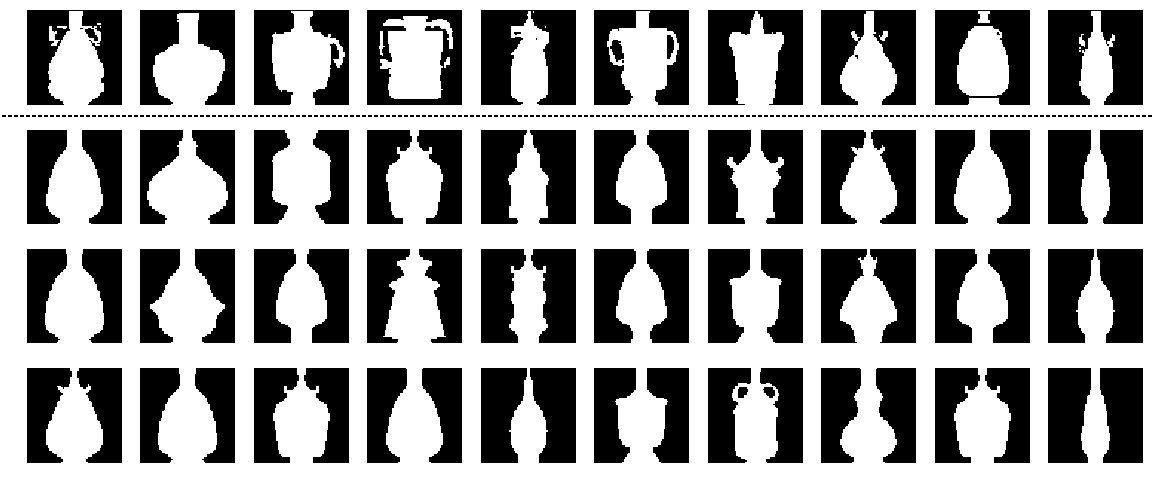}
	\caption{A gallery of 10 randomly sampled vases generated by \OurNet{} (top row) and below, their 3-nearest-neighbors from the training set, based on pixel-wise Euclidean distance. One can observe that the generated vases are different from their nearest neighbors.}
	\label{fig:vase_nn_train}
\end{figure}

\begin{figure*}
	\centering
		\begin{minipage}{0.47\linewidth}
     \centering
	\includegraphics[width=\linewidth]{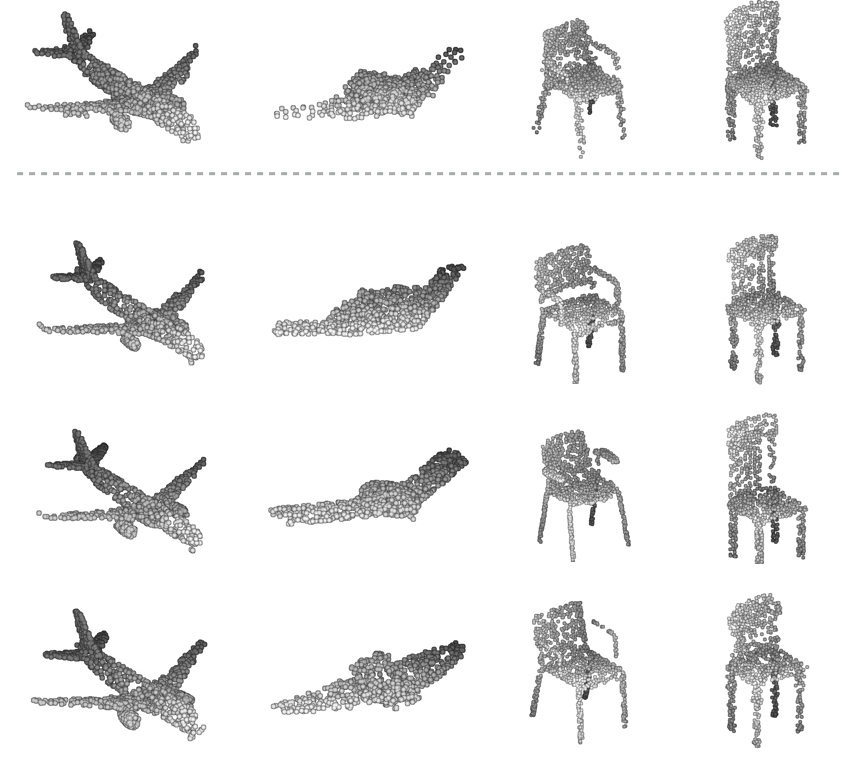}
	\centerline{baseline~\cite{achlioptas2018learning}}
   \end{minipage}\hfill
   		\begin{minipage}{0.47\linewidth}
     \centering
		\includegraphics[width=\linewidth]{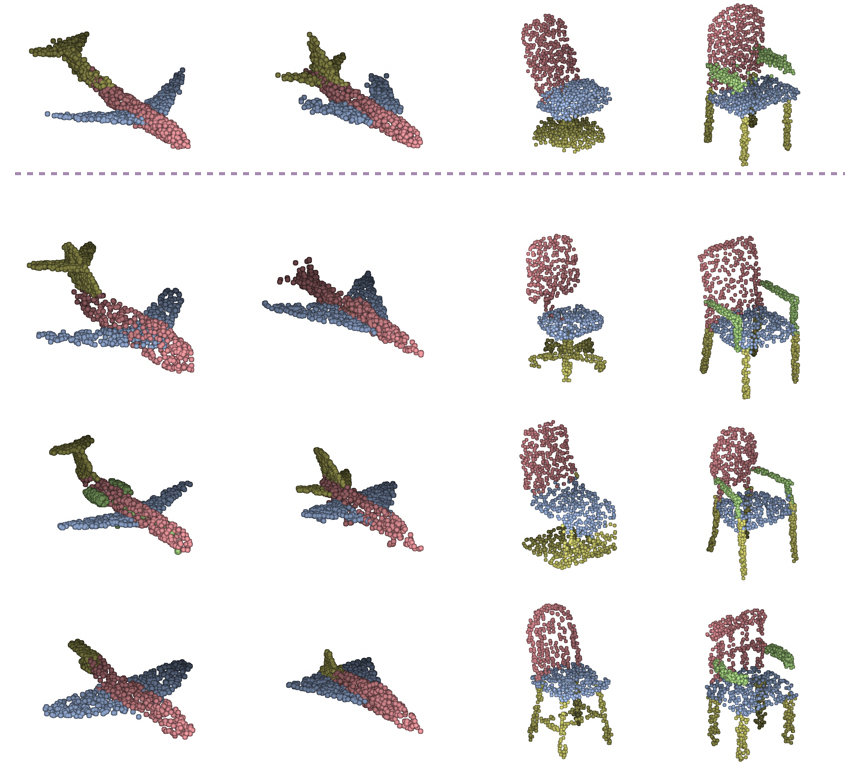}
		\centerline{Ours}
   \end{minipage}\hfill
	\caption{Qualitative comparison of diversity. The generated data of both methods (first row) is realistic. However, searching for the nearest neighbors of the generated data in the training set (rows two to four) reveals that our method (right side) exhibits more diversity compared to the baseline~\cite{achlioptas2018learning} (left side). Please note that the baseline's shapes are presented in gray to emphasize that the baseline generates the entire shape, which is not segmented to semantic parts.}
	\label{fig:3d_nn_train}
\end{figure*}

\subsection{Qualitative evaluation}
\label{sec: Qualitative evaluation}

We evaluate our network, \OurNet{}, on 2D data and 3D point clouds. Figure~\ref{fig:gallery_results} shows some generated 3D results. Unlike other naive approaches, we are able to generate versatile samples beyond the empirical distribution. In order to visualize the versatility, we present the nearest neighbor of the generated samples in the training set. As shown in Figure~\ref{fig:vase_nn_train}, for the 2D case, samples generated by our generative approach differ from the closest training samples. In Figure~\ref{fig:3d_nn_train} we also compare this qualitative diversity measure with the baseline~\cite{achlioptas2018learning}, showing that our generated samples are more distinct from their nearest neighbors in the training set, compared to those generated by the baseline. In the following sections, we quantify this attribute. More generated results can be found in the supplementary material.

\subsection{Quantitative evaluation}
We quantify the ability of our model to generate realistic unseen samples using two novel metrics. To evaluate our model, we use 5,000 randomly sampled shapes from our trained model and from the baselines.

\vspace{-10pt}

\begin{figure}
	\centering
    \includegraphics[width=0.6\linewidth]{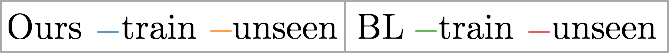}
	\includegraphics[width=\linewidth]{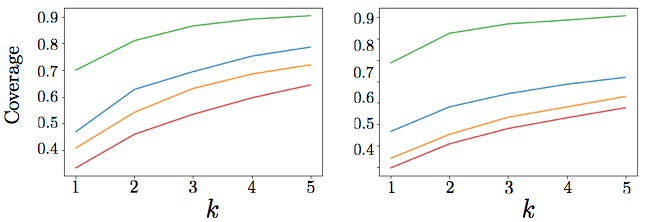}
	\caption{$k$-set-overage comparison for chairs (left) and airplanes (right) point clouds sets. We generate an equal number of samples in both our method and the baseline \cite{achlioptas2018learning} and calculate the $k$-set-coverage of both the training set and unseen set by them. While the baseline covers the training set almost perfectly, it has lower coverage of the unseen set. Our method \OurNet{} balances between generating samples similar to the training set and the unseen set.}
	\label{fig:set_coverage}
\end{figure}

\paragraph{$k$-set-coverage.}
We define the $k$-set-coverage of set $A$ by set $B$ as the percentage $P$ of shapes from $A$ which are one of the $k$-nearest-neighbors of some shape in $B$. Thus, if the set $B$ is similar only to a small part of set $A$, the $k$-Set-Coverage will be small and vice-versa.
In our case, we calculate the nearest neighbors using Chamfer distance. In Figure~\ref{fig:set_coverage}, we compare the $k$-set-coverage of the unseen set and the training set by our generated data and by the baseline~\cite{achlioptas2018learning} generated data. It is clear that the baseline covers the training better, since most of its samples lie close to it. However, the unseen set is covered poorly by the baseline, for all $k$, while, our method, balances between generating seen samples and unseen samples.

\vspace{-10pt}

\paragraph{Diversity.}
We develop a second measure to quantify the generated unseen data, which relies on a trained classifier to distinguish between the training and the unseen set. Then, we measure the percentage of generated shapes which are classified as belonging to the unseen set. The classifier architecture is a straight-forward adaption of the encoder from the part synthesis unit of the training process, followed by fully connected layers which classify between unseen and train sets (see supplementary file for details).

Table~\ref{table:classifier_res} shows some classification results for generated vases, chairs, and airplanes by our method and two baselines. We can observe that when the seen set is relatively small, e.g., $5\%$ or $15\%$ of the total, our model clearly performs better than the baselines in terms of generative diversity, as exhibited by the higher levels of coverage over the unseen set. However, as the seen set increases in size, e.g., to $30\%$, the difference between our method and the baselines becomes smaller. We believe that this trend is not a reflection that our method starts to generate less diverse samples, but rather that the unseen set is becoming more similar to the seen set, hence less diverse itself. %

\begin{figure}
\centering
    \begin{minipage}{1.0\linewidth}
    \begin{minipage}{0.04\linewidth}
	 \rotatebox[origin=c]{90}{Airplanes}
	 \end{minipage}
	 \centering
	\begin{minipage}{0.95\linewidth}
     \centering
     \includegraphics[width=\linewidth]{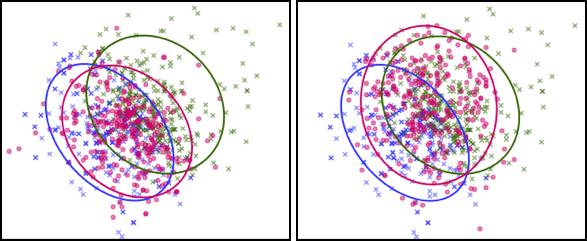}
   \end{minipage}\hfill
   \end{minipage}
   \vspace{0.1cm}
   
    \begin{minipage}{1.0\linewidth}
    \begin{minipage}{0.04\linewidth}
	 \rotatebox[origin=c]{90}{Chairs}
	 \end{minipage}
	 \centering
	\begin{minipage}{0.95\linewidth}
     \centering
     \includegraphics[width=\linewidth]{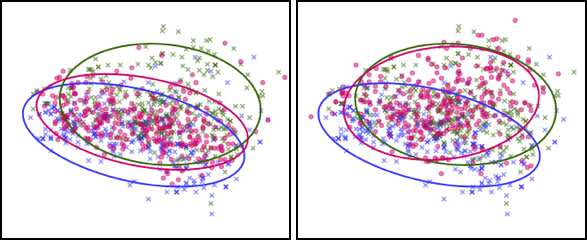}
   \end{minipage}\hfill
   \end{minipage}
   
\noindent \phantom{-------------} Baseline \cite{achlioptas2018learning} \hfill Ours \phantom{--------------}
	\caption{A visualization of the classifiers' feature spaces for 3D data. The classifier is trained to distinguish between the training set (purple crosses) and the unseen set (green crosses). The two sets are clearly visible in the resulting space. While the baseline method \cite{achlioptas2018learning} (pink dots) generates samples similar to the seen set, \OurNet{} (pink dots) generates samples in the unseen region.}
	\label{fig:embedding}
\end{figure}

To visualize the coverage of seen/unseen regions by the generated samples, we use the classifier's embedding (the layer before the final fully-connected layer) and reduce its dimension by projecting it onto the 2D PCA plane, as shown in Figures~\ref{fig:teaser} and ~\ref{fig:embedding}. The training and unseen sets have overlap in this representation, reflecting data which is similar between the two sets.  While both methods are able to generate unseen samples in the overlap region, the baseline samples are biased toward the training set. In contrary, our generated samples are closer to the unseen.

\begin{table}[!t]
    \begin{subtable}{0.48\textwidth}
	\centering
    	\begin{tabular}[c]{| c | c | c | c | }
    		\hline		
    		Category & Vase & Chair & Airplane \\
            \hhline{|=|=|=|=|}	
    		VAE &  0.3$\pm$0.08 & - & - \\
    		\cline{1-4}
    		WGAN-GP & - & 0.67$\pm$0.02 & 0.66$\pm$0.03 \\
    		\cline{1-4}
    		AE+GMM~\cite{achlioptas2018learning} & - & 0.61$\pm$0.03 & 0.67$\pm$0.12 \\
    		\cline{1-4}
    		Ours & \bf{0.46$\pm$0.08} & \bf{0.76$\pm$0.06} & \bf{0.8$\pm$0.02} \\
    		\hline  
    	\end{tabular}
    	\vspace{-2pt}
    	\caption{\footnotesize{2D vase: $50\%$-$50\%$ (seen-unseen); 3D: $15\%$-$85\%$ (seen-unseen)}}
        \label{tab:table1_a}
	\end{subtable}

	\vspace{3pt}

	\begin{subtable}{0.48\textwidth}
	\centering
    	\begin{tabular}[c]{| c | c | c | c | }
    		\hline		
    		Category & WGAN-GP & AE+GMM~\cite{achlioptas2018learning} & Ours\\
            \hhline{|=|=|=|=|}	
    		Chair & 0.75$\pm$0.05 & 0.57$\pm$0.06 & \bf{0.87$\pm$0.05} \\
    		\cline{1-4}
    		Airplane & 0.76$\pm$0.07 & 0.43$\pm$0.1 & \bf{0.86$\pm$0.04} \\
    		\hline  
    	\end{tabular}
    	\vspace{-2pt}
    	\caption{\footnotesize{$5\%$-$95\%$ (seen-unseen)}}
        \label{tab:table1_b}
	\end{subtable}

	\vspace{3pt}

	\begin{subtable}{0.48\textwidth}
	\centering
    	\begin{tabular}[c]{| c | c | c | c | }
    		\hline		
    		Category & WGAN-GP & AE+GMM~\cite{achlioptas2018learning} & Ours\\
            \hhline{|=|=|=|=|}	
    		Chair & 0.54$\pm$0.07 & \bf{0.63$\pm$0.07} & 0.55$\pm$0.01 \\
    		\cline{1-4}
    		Airplane & 0.61$\pm$0.07 & 0.52$\pm$0.12 & \bf{0.65$\pm$0.1} \\
    		\hline  
    	\end{tabular}
    	\vspace{-2pt}
    	\caption{\footnotesize{$30\%$-$70\%$ (seen-unseen)}}
        \label{tab:table1_c}
	\end{subtable}
	\caption{Comparing generative diversity between baselines and \OurNet{}. We report the percentage of generated samples that are classified as belonging to the unseen set, averaged over five random splits, for three split percentages. }
	\label{table:classifier_res}
\end{table}

\paragraph{JSD.}
The Jensen-Shannon divergence is a distance measure between two probability distributions and is given by
\vspace{-5pt}
\begin{equation}
    \textrm{JSD}\left(S || T\right) = \frac{1}{2}D\left(S||M\right) + \frac{1}{2}D\left(T||M\right),
\vspace{-5pt}
\end{equation}
where $S$, $T$ are probability distributions, $M=\frac{1}{2}\left(S+T\right)$ and $D$ is the KL-divergence \cite{kullback1951}.
Following~\cite{achlioptas2018learning} we define the occupancy probability distribution for a set of point clouds by counting the number of points lying within each voxel in a regular voxel grid. Assuming the point clouds are normalized and axis-aligned, the JSD between two such probabilities measure the degree to which two point cloud sets occupy similar locations. Thus, we calculate the occupancy matrix for the unseen set and compare it to the occupancy matrix of samples generated by our method and the baselines. The results are summarized in Table~\ref{table:jsd} and clearly show that our generated samples are closer to the unseen set structure. The number of voxels used is $28^3$, as in ~\cite{achlioptas2018learning} and the size of all point clouds sets is equal.

\begin{table}[!t]
	\centering
	\begin{tabular}[c]{| c | c | c | c | }
		\hline		
		Category & WGAN-GP & AE+GMM~\cite{achlioptas2018learning} & Ours \\
        \hhline{|=|=|=|=|}	
		Chair & 0.3$\pm$0.02 & 0.19$\pm$0.006 &  \bf{0.02$\pm$0.003} \\
		\cline{1-4}
		Airplanes & 0.32$\pm$0.016 & 0.14$\pm$0.007 &  \bf{0.07$\pm$0.013} \\
		\hline  
	\end{tabular}
	\caption{The JSD distance between the unseen set and the generated samples from \OurNet{} and the baselines, averaged over five random seen-unseen splits.}
	\label{table:jsd}
\end{table}

%% file: future.tex
\section{Conclusion, limitation, and future work}
 \label{sec:future}

We believe that effective generative models should strive to venture more into the ``unseen'' data of a target distribution, beyond the observed exemplars from the training set. Covering both the seen and the unseen implies that the generated data is both {\em fit\/} and {\em diverse\/}~\cite{xu2012fit}. Fitness constrains the generated data to be close to data from the target domain, both the seen and the unseen. Diversity ensures that the generated data is not confined only to the seen data.

We have presented a generic approach for ``fit-n-diverse'' shape modeling based on a part-based prior, where a shape is not viewed as an unstructured whole but as the result of a coherent composition of a set of parts. This is realized by \OurNet{}, a novel deep generative network composed of a part synthesis unit and a part composition unit. Novel shapes are generated via inference over random samples taken from the latent spaces of shape parts and part compositions. Our work also contributes two novel measures to evaluate generative models: the $k$-set-coverage and a diversity measure which quantifies the percentage of generated data classified as ``unseen" vs.~data from the training set.

Compared to baseline approaches, our generative network demonstrates superiority, but still somewhat limited diversity, since the generative power of the part-based approach is far from being fully realized. Foremost, an intrinsic limitation of our composition mechanism is that it is still ``in place'': it does not allow changes to part structures or feature transfers between different part classes. For example, enabling a simple symmetric switch in part composition would allow the generation of right hand images when all the training images are of the left hand.

\OurNet{} can be directly applied for generative modeling of organic shapes. But in terms of plausibility, such shapes place a more stringent requirement on coherent and smooth part connections, an issue that our current method does not account for. Perfecting part connections can be a post process. Learning a deep model for the task is worth pursuing for future work. Our current method is also limited by the spatial transformations allowed by the STN during part composition. As a result, we can only deal with man-made shapes without part articulation.

As more immediate future work, we would like to apply our approach to more complex datasets, where parts can be defined during learning. In general, we believe that more research shall focus on other generation related prior information, besides parts-based priors. %
Further down the line, we envision that the fit-n-diverse approach, with generative diversity, will form a baseline for {\em creative modeling}~\cite{cohenor2016}, potentially allowing part exchanges across different object categories. This may have to involve certain perceptual studies or scores, to judge creativity. The compelling challenge is how to define a generative neural network with sufficient diversity to cross the line of being creative~\cite{DBLP:journals/corr/ElgammalLEM17}.

%% file: supp.tex
\section{Supplementary Material}

\subsection{Network Architectures}
\label{sec:net_arc}
\paragraph{Part synthesis}
The architectures of the part synthesis generative autoencoders, for both 3D and 2D cases, are listed in Table~\ref{tab:part_syn_3d} and Table~\ref{tab:part_syn_2d} respectively. We used the following standard hyper-parameters to train the 3D (2D) model: Adam optimizer, $\beta_1=0.9 (0.5)$, $\beta_2=0.999$, learning rate $= 0.001 (2e^{-4})$, batch size $= 64$.

\begin{table}[h!]
	\begin{small}
		\begin{tabular}{lllll} \hline
			Operation & Kernel & Strides & Feature maps & Act. func. \\ \hline\hline
			\multicolumn{5}{l}{Encode $x$: 400x3 point cloud $\rightarrow$ 64-dim feature vector $f_x$} \\ \hline
			1D conv. & 1x64  & 1 & 400x64 & Relu \\
			1D conv. & 1x64  & 1 & 400x64 & Relu \\
			1D conv. & 1x64  & 1 & 400x64 & Relu \\
            1D conv. & 1x128  & 1 & 400x128 & Relu \\
            1D conv. & 1x64  & 1 & 400x64 & Relu \\
            Max Pooling & -- & -- & 128 & -- \\ \hline
			\multicolumn{5}{l}{Decode $f_x$: 64-dim feature vector $\rightarrow$ 400x3 point cloud $x'$} \\ \hline
			Linear & -- & -- & 256 & Relu \\
			Linear & -- & -- & 256 & Relu \\ 
            Linear & -- & -- & 400x3 & -- \\ \hline
		\end{tabular}
	\end{small}
	\caption{Layers of the 3D part generative AE shown in order from input to output.}
	\label{tab:part_syn_3d}
\end{table}

\begin{table}[h!]
	\begin{small}
		\begin{tabular}{lllll} \hline
			Operation & Kernel & Strides & Feature maps & Act. func. \\ \hline\hline
			\multicolumn{5}{l}{Encode $x$: 64x64x1 input shape $\rightarrow$ 10-dim feature vector $f_x$} \\ \hline
			Conv. & 5x5x8  & 2x2 & 32x32x8 & l-Relu \\
            Conv. & 5x5x16  & 2x2 & 16x16x16 & l-Relu \\
            Conv. & 5x5x32  & 2x2 & 8x8x32 & l-Relu \\
            Conv. & 5x5x64  & 2x2 & 4x4x64 & l-Relu \\
			2xLinear & -- & -- & 10 & -- \\ \hline
			\multicolumn{5}{l}{Decode $f_x$: 10-dim feature vector $\rightarrow$ 64x64x1 shape $x$} \\ \hline
            Linear & -- & -- & 1024 (=4x4x64) & Relu \\
			Trans. conv. & 5x5x32 & 2x2 & 8x8x32 & Relu \\
			Trans. conv. & 5x5x16 & 2x2 & 16x16x16 & Relu \\
            Trans. conv. & 5x5x8 & 2x2 & 32x32x8 & Relu \\
            Trans. conv. & 5x5x1 & 2x2 & 64x64x1 & Sigmoid \\\hline
		\end{tabular}
	\end{small}
	\caption{Layers of the 2D part VAE shown in order from input to output. After the last conv. layer we have two parallel linear layers for computing the mean and std of the VAE.}
	\label{tab:part_syn_2d}
\end{table}

\paragraph{Parts composition}
The architectures of the parts composition units are listed in Table~\ref{tab:parts_comp_3d} and Table~\ref{tab:parts_comp_2d}, for the 3D and 2D cases respectively. We used the following standard hyper-parameters to train the 3D and 2D models: Adam optimizer, $\beta_1=0.9$, $\beta_2=0.999$, learning rate $= 0.001$, batch size $= 64$.

\begin{table}[h!]
	\begin{small}
		\begin{tabular}{lll} \hline
			Operation & Feature maps & Act. func. \\ \hline\hline
			\multicolumn{3}{l}{Comp. net $x$: 64xC+16 feature vector $\rightarrow$ 6XC comp. vector $f_x$} \\ \hline
			Linear & 256 & Relu \\
			Linear & 128 & Relu \\ 
            Linear & 6xC & -- \\ \hline\\
		\end{tabular}
	\end{small}
	\caption{Layers of the 3D parts composition network shown in order from input to output. Where $C$ is the number of parts. The input vector is a concatenation of $C$ feature vectors of size $64$ with a noise vector of size $16$.}
	\label{tab:parts_comp_3d}
\end{table}

\begin{table}[h!]
	\begin{small}
		\begin{tabular}{lll} \hline
			Operation & Feature maps & Act. func. \\ \hline\hline
			\multicolumn{3}{l}{Comp. net $x$: 10xC+8 feature vector $\rightarrow$ 4XC comp. vector $f_x$} \\ \hline
			Linear & 128 & Relu \\
			Linear & 128 & Relu \\ 
            Linear & 4xC & -- \\ \hline\\
		\end{tabular}
	\end{small}
	\caption{Layers of the 2D parts composition network shown in order from input to output. Where $C$ is the number of parts. The input vector is a concatenation of $C$ feature vectors of size $10$ with a noise vector of size $8$.}
	\label{tab:parts_comp_2d}
\end{table}

\subsection{More results}
\label{sec:more_results}
In this section, we present additional results of \OurNet{} for the three categories. For the Chair and Airplane categories we have randomly sampled $80$ shapes from the $5,000$ we have generated for the quantitative metrics; see Figure~\ref{fig:supp_chair_gallery} and Figure~\ref{fig:supp_plane_gallery} respectively. For the Vases, since the output is smaller, we sampled $300$ shapes from the $1,024$ we generated for the quantitative metrics; see Figure~\ref{fig:supp_vase_gallery}.
Furthermore, we present additional interpolation results from \OurNet{} on the 3D categories; see Figure~\ref{fig:supp_interpolation_chair_full} and Figure~\ref{fig:supp_interpolation_plane_full} for linear interpolations, Figure~\ref{fig:supp_interpolation_chair_part} and Figure~\ref{fig:supp_interpolation_plane_part} for part-by-part interpolations.

\subsection{Comparison to baseline}
\label{sec:baseline}
In this section, we present randomly picked generated results from \OurNet{} and the baseline on the Chair and Airplane categories. For each generated shape, we present its three nearest neighbors based on the Chamfer distance. We present the results side by side to emphasize the power of \OurNet{} in generating novel shapes in comparison to the baseline; see Figure~\ref{fig:chairs_nn} for chairs and Figure~\ref{fig:plane_nn} for airplanes.

\include{chairs_gallery}
\include{plane_gallery}
\include{vase_gallery}
\include{supp_interpolation_chair}
\include{supp_interpolation_plane}
\include{chairs_nn}
\include{plane_nn}

%% file: chairs_gallery.tex
\begin{figure*}
	\centering
\begin{minipage}{\linewidth}\centering
\includegraphics[width=0.12\linewidth]{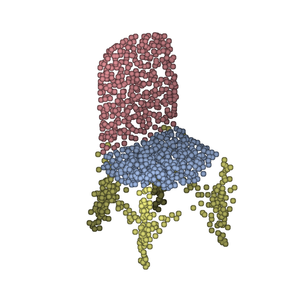}
\includegraphics[width=0.12\linewidth]{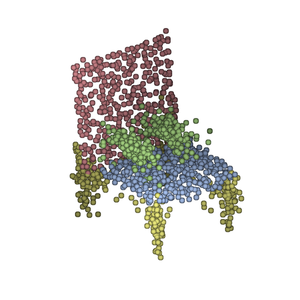}
\includegraphics[width=0.12\linewidth]{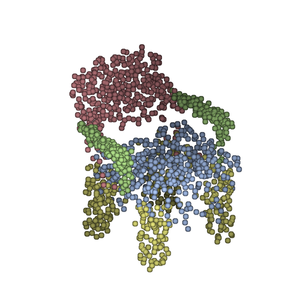}
\includegraphics[width=0.12\linewidth]{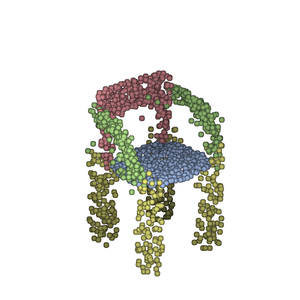}
\includegraphics[width=0.12\linewidth]{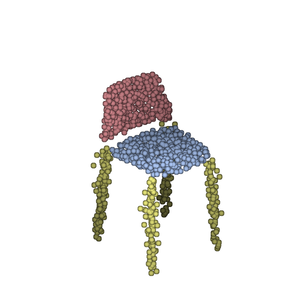}
\includegraphics[width=0.12\linewidth]{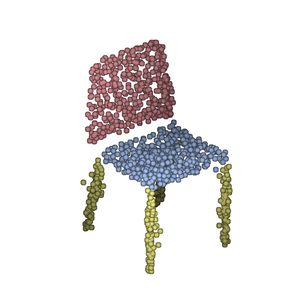}
\includegraphics[width=0.12\linewidth]{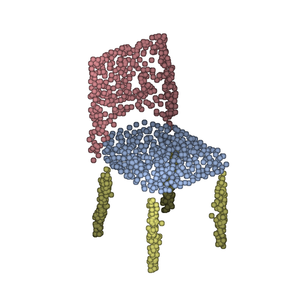}
\includegraphics[width=0.12\linewidth]{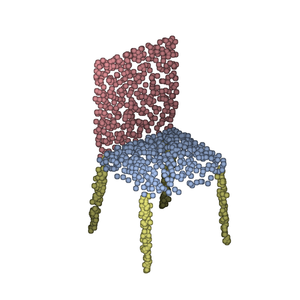}
\end{minipage}

\begin{minipage}{\linewidth}\centering
\includegraphics[width=0.12\linewidth]{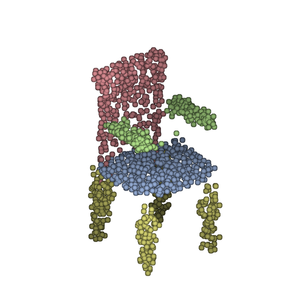}
\includegraphics[width=0.12\linewidth]{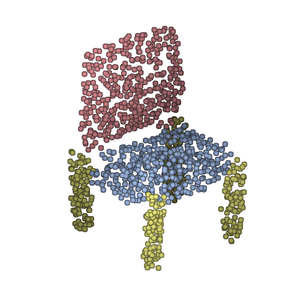}
\includegraphics[width=0.12\linewidth]{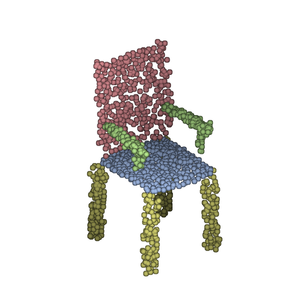}
\includegraphics[width=0.12\linewidth]{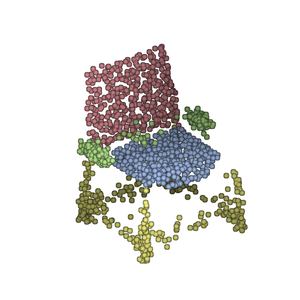}
\includegraphics[width=0.12\linewidth]{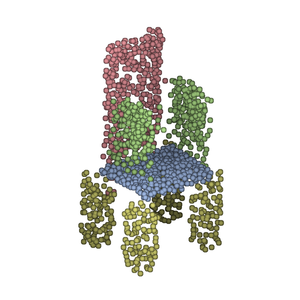}
\includegraphics[width=0.12\linewidth]{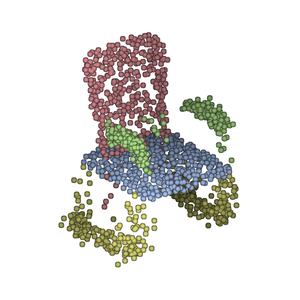}
\includegraphics[width=0.12\linewidth]{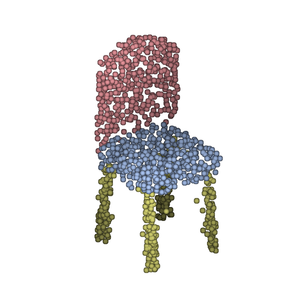}
\includegraphics[width=0.12\linewidth]{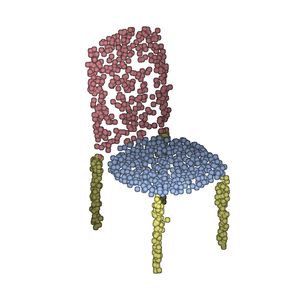}
\end{minipage}

\begin{minipage}{\linewidth}\centering
\includegraphics[width=0.12\linewidth]{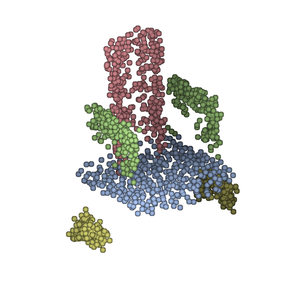}
\includegraphics[width=0.12\linewidth]{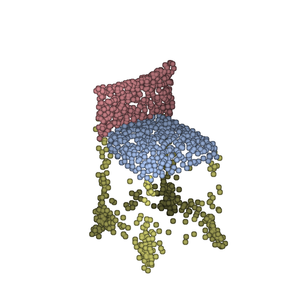}
\includegraphics[width=0.12\linewidth]{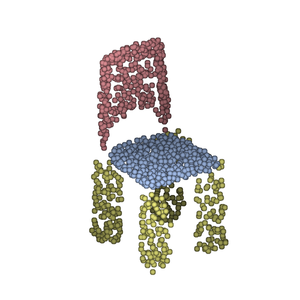}
\includegraphics[width=0.12\linewidth]{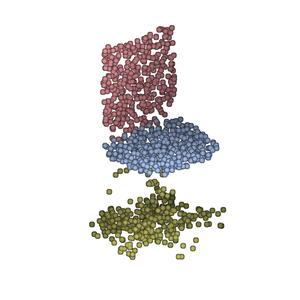}
\includegraphics[width=0.12\linewidth]{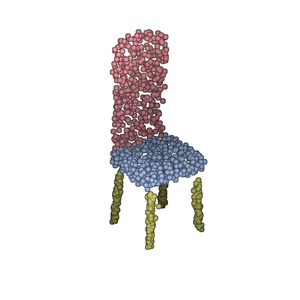}
\includegraphics[width=0.12\linewidth]{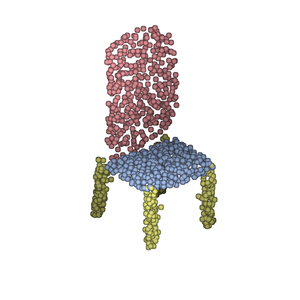}
\includegraphics[width=0.12\linewidth]{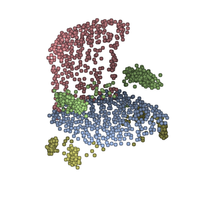}
\includegraphics[width=0.12\linewidth]{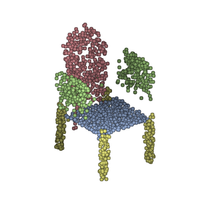}
\end{minipage}

\begin{minipage}{\linewidth}\centering
\includegraphics[width=0.12\linewidth]{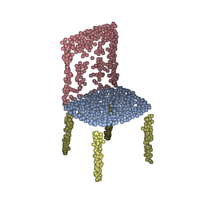}
\includegraphics[width=0.12\linewidth]{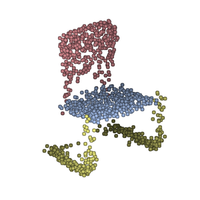}
\includegraphics[width=0.12\linewidth]{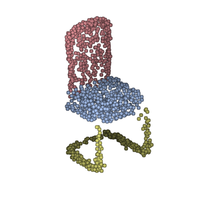}
\includegraphics[width=0.12\linewidth]{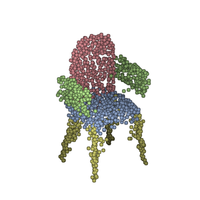}
\includegraphics[width=0.12\linewidth]{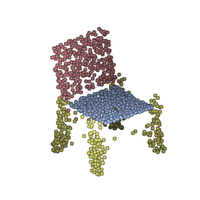}
\includegraphics[width=0.12\linewidth]{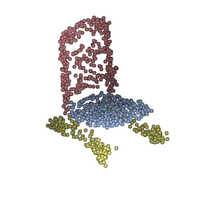}
\includegraphics[width=0.12\linewidth]{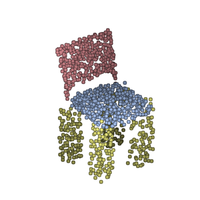}
\includegraphics[width=0.12\linewidth]{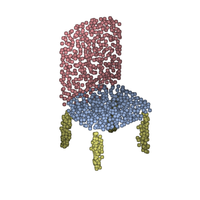}
\end{minipage}

\begin{minipage}{\linewidth}\centering
\includegraphics[width=0.12\linewidth]{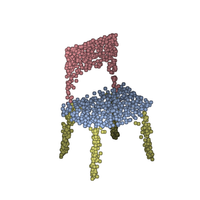}
\includegraphics[width=0.12\linewidth]{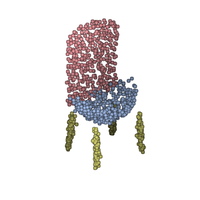}
\includegraphics[width=0.12\linewidth]{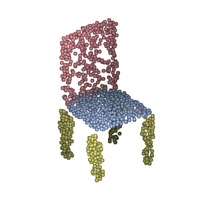}
\includegraphics[width=0.12\linewidth]{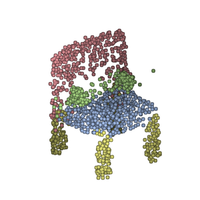}
\includegraphics[width=0.12\linewidth]{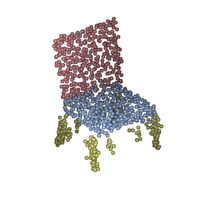}
\includegraphics[width=0.12\linewidth]{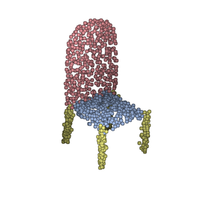}
\includegraphics[width=0.12\linewidth]{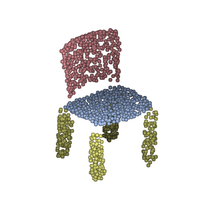}
\includegraphics[width=0.12\linewidth]{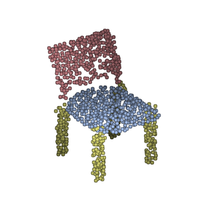}
\end{minipage}

\begin{minipage}{\linewidth}\centering
\includegraphics[width=0.12\linewidth]{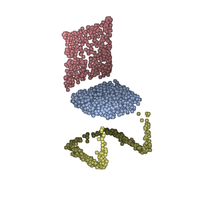}
\includegraphics[width=0.12\linewidth]{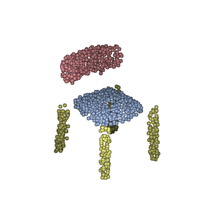}
\includegraphics[width=0.12\linewidth]{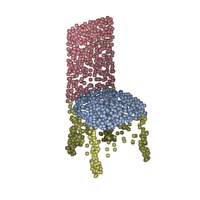}
\includegraphics[width=0.12\linewidth]{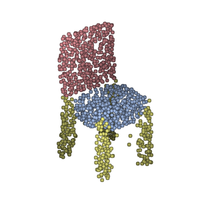}
\includegraphics[width=0.12\linewidth]{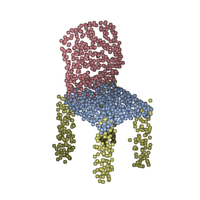}
\includegraphics[width=0.12\linewidth]{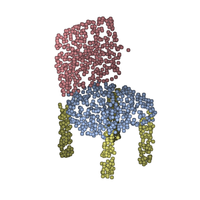}
\includegraphics[width=0.12\linewidth]{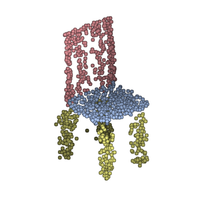}
\includegraphics[width=0.12\linewidth]{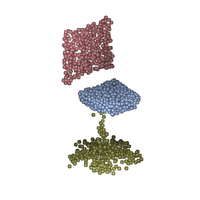}
\end{minipage}

\begin{minipage}{\linewidth}\centering
\includegraphics[width=0.12\linewidth]{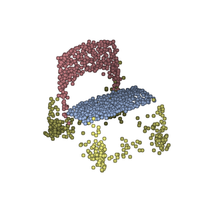}
\includegraphics[width=0.12\linewidth]{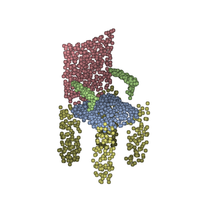}
\includegraphics[width=0.12\linewidth]{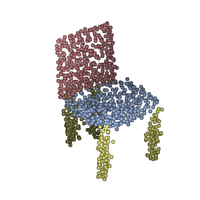}
\includegraphics[width=0.12\linewidth]{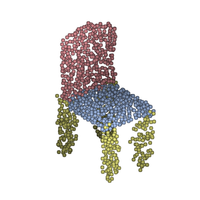}
\includegraphics[width=0.12\linewidth]{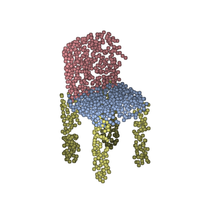}
\includegraphics[width=0.12\linewidth]{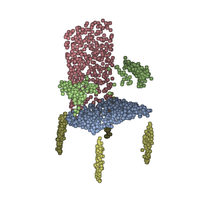}
\includegraphics[width=0.12\linewidth]{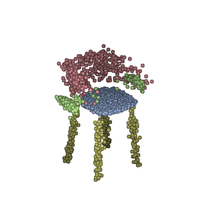}
\includegraphics[width=0.12\linewidth]{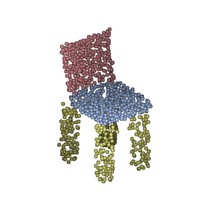}
\end{minipage}

\begin{minipage}{\linewidth}\centering
\includegraphics[width=0.12\linewidth]{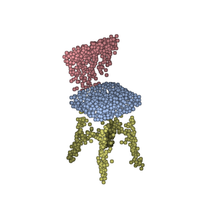}
\includegraphics[width=0.12\linewidth]{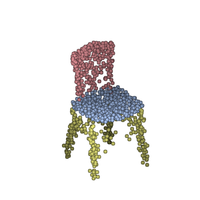}
\includegraphics[width=0.12\linewidth]{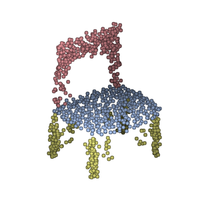}
\includegraphics[width=0.12\linewidth]{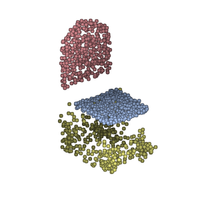}
\includegraphics[width=0.12\linewidth]{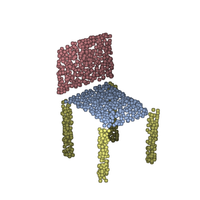}
\includegraphics[width=0.12\linewidth]{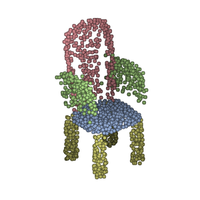}
\includegraphics[width=0.12\linewidth]{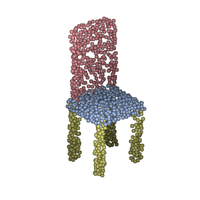}
\includegraphics[width=0.12\linewidth]{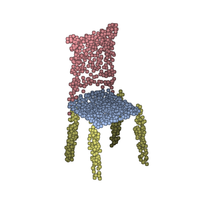}
\end{minipage}

\begin{minipage}{\linewidth}\centering
\includegraphics[width=0.12\linewidth]{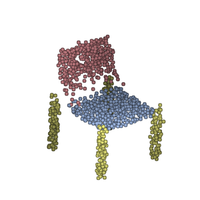}
\includegraphics[width=0.12\linewidth]{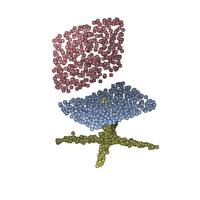}
\includegraphics[width=0.12\linewidth]{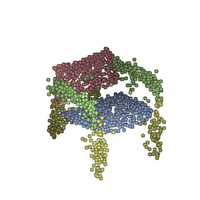}
\includegraphics[width=0.12\linewidth]{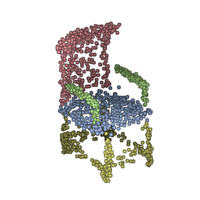}
\includegraphics[width=0.12\linewidth]{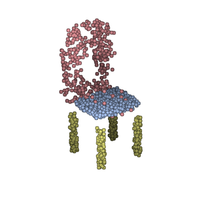}
\includegraphics[width=0.12\linewidth]{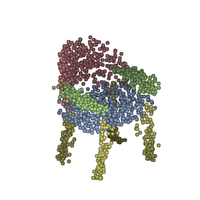}
\includegraphics[width=0.12\linewidth]{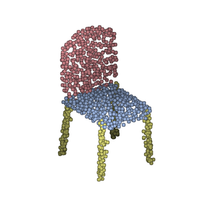}
\includegraphics[width=0.12\linewidth]{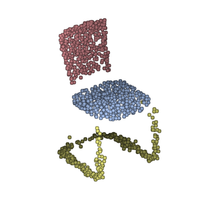}
\end{minipage}

\begin{minipage}{\linewidth}\centering
\includegraphics[width=0.12\linewidth]{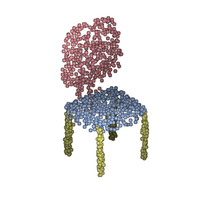}
\includegraphics[width=0.12\linewidth]{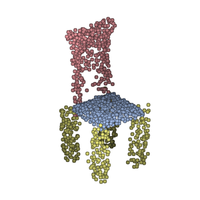}
\includegraphics[width=0.12\linewidth]{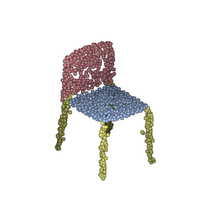}
\includegraphics[width=0.12\linewidth]{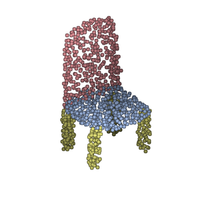}
\includegraphics[width=0.12\linewidth]{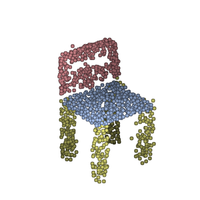}
\includegraphics[width=0.12\linewidth]{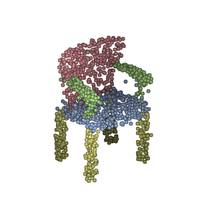}
\includegraphics[width=0.12\linewidth]{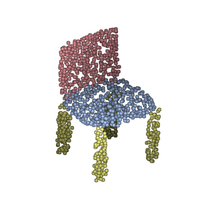}
\includegraphics[width=0.12\linewidth]{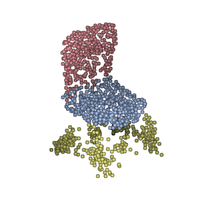}
\end{minipage}

	\caption{Chair gallery. We present $80$ randomly sampled chairs from the $5,000$ which were generated by \OurNet{}.}
	\label{fig:supp_chair_gallery}
\end{figure*}

%% file: plane_gallery.tex
\begin{figure*}
	\centering
\begin{minipage}{\linewidth}\centering
\includegraphics[width=0.12\linewidth]{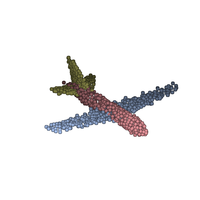}
\includegraphics[width=0.12\linewidth]{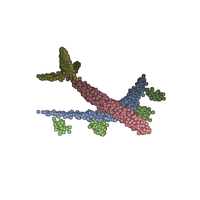}
\includegraphics[width=0.12\linewidth]{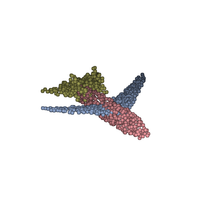}
\includegraphics[width=0.12\linewidth]{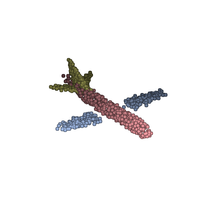}
\includegraphics[width=0.12\linewidth]{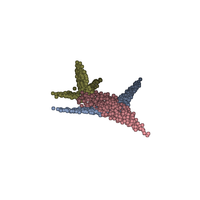}
\includegraphics[width=0.12\linewidth]{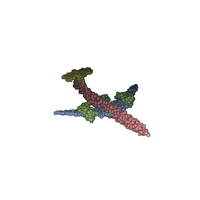}
\includegraphics[width=0.12\linewidth]{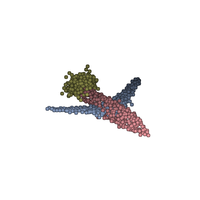}
\includegraphics[width=0.12\linewidth]{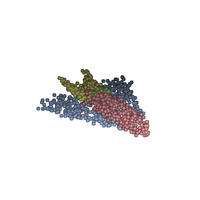}
\end{minipage}

\begin{minipage}{\linewidth}\centering
\includegraphics[width=0.12\linewidth]{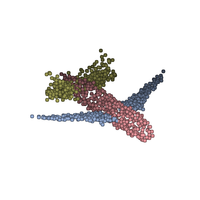}
\includegraphics[width=0.12\linewidth]{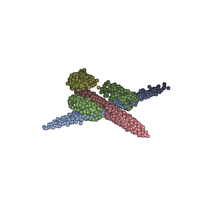}
\includegraphics[width=0.12\linewidth]{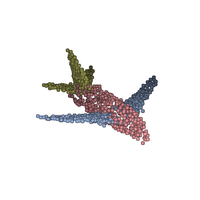}
\includegraphics[width=0.12\linewidth]{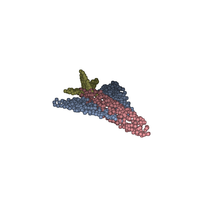}
\includegraphics[width=0.12\linewidth]{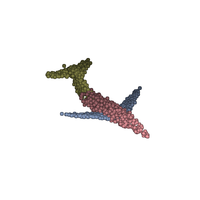}
\includegraphics[width=0.12\linewidth]{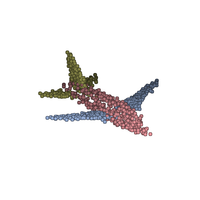}
\includegraphics[width=0.12\linewidth]{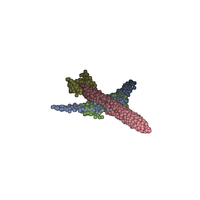}
\includegraphics[width=0.12\linewidth]{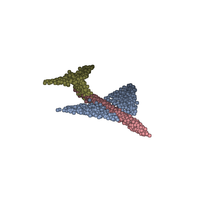}
\end{minipage}

\begin{minipage}{\linewidth}\centering
\includegraphics[width=0.12\linewidth]{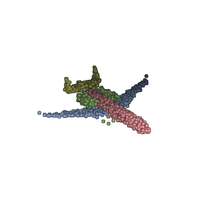}
\includegraphics[width=0.12\linewidth]{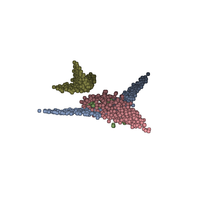}
\includegraphics[width=0.12\linewidth]{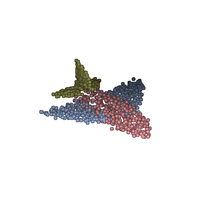}
\includegraphics[width=0.12\linewidth]{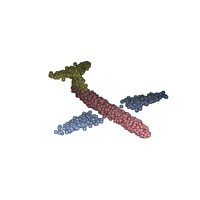}
\includegraphics[width=0.12\linewidth]{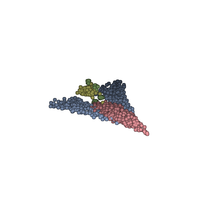}
\includegraphics[width=0.12\linewidth]{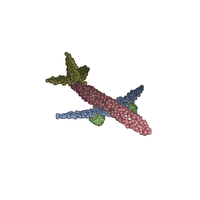}
\includegraphics[width=0.12\linewidth]{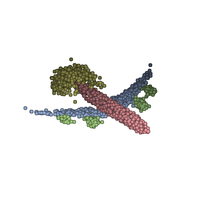}
\includegraphics[width=0.12\linewidth]{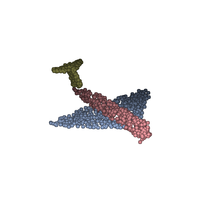}
\end{minipage}

\begin{minipage}{\linewidth}\centering
\includegraphics[width=0.12\linewidth]{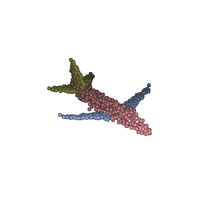}
\includegraphics[width=0.12\linewidth]{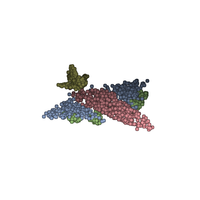}
\includegraphics[width=0.12\linewidth]{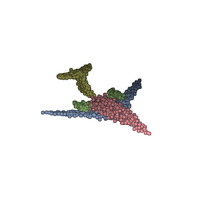}
\includegraphics[width=0.12\linewidth]{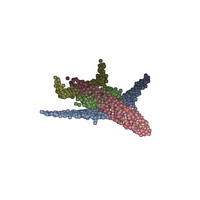}
\includegraphics[width=0.12\linewidth]{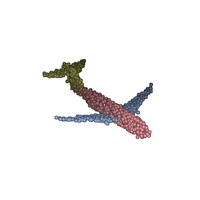}
\includegraphics[width=0.12\linewidth]{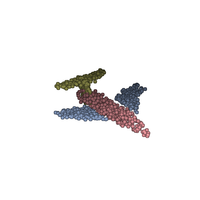}
\includegraphics[width=0.12\linewidth]{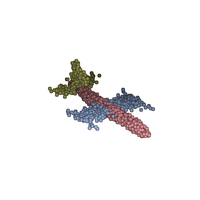}
\includegraphics[width=0.12\linewidth]{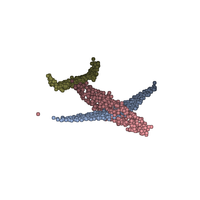}
\end{minipage}

\begin{minipage}{\linewidth}\centering
\includegraphics[width=0.12\linewidth]{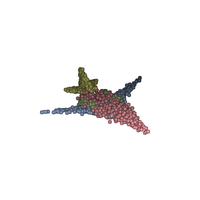}
\includegraphics[width=0.12\linewidth]{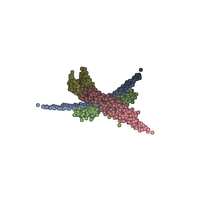}
\includegraphics[width=0.12\linewidth]{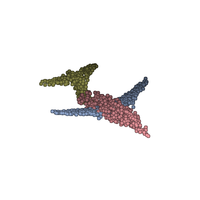}
\includegraphics[width=0.12\linewidth]{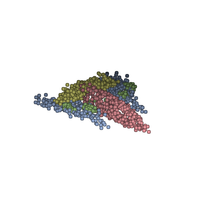}
\includegraphics[width=0.12\linewidth]{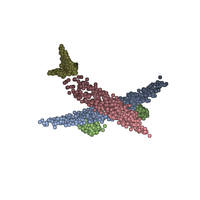}
\includegraphics[width=0.12\linewidth]{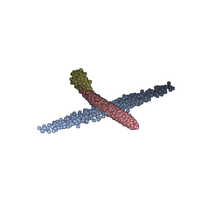}
\includegraphics[width=0.12\linewidth]{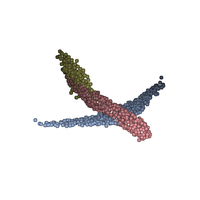}
\includegraphics[width=0.12\linewidth]{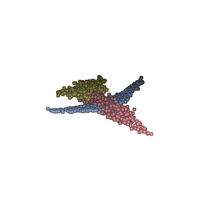}
\end{minipage}

\begin{minipage}{\linewidth}\centering
\includegraphics[width=0.12\linewidth]{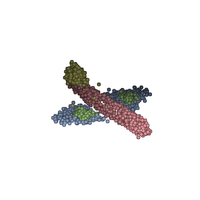}
\includegraphics[width=0.12\linewidth]{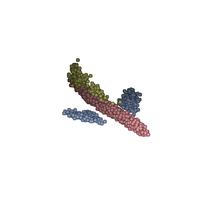}
\includegraphics[width=0.12\linewidth]{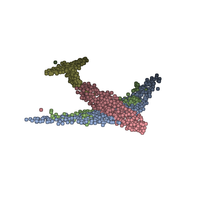}
\includegraphics[width=0.12\linewidth]{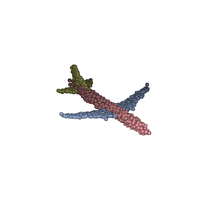}
\includegraphics[width=0.12\linewidth]{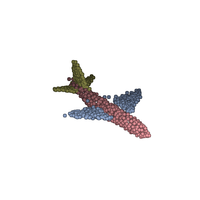}
\includegraphics[width=0.12\linewidth]{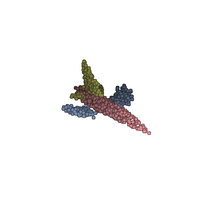}
\includegraphics[width=0.12\linewidth]{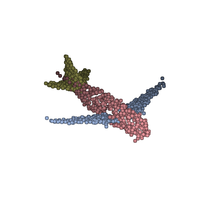}
\includegraphics[width=0.12\linewidth]{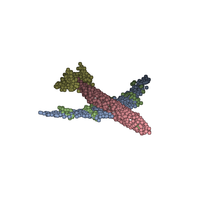}
\end{minipage}

\begin{minipage}{\linewidth}\centering
\includegraphics[width=0.12\linewidth]{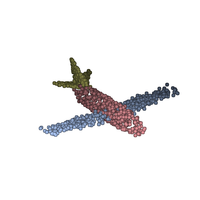}
\includegraphics[width=0.12\linewidth]{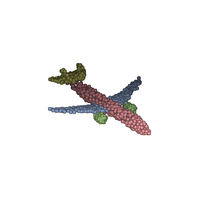}
\includegraphics[width=0.12\linewidth]{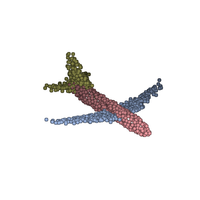}
\includegraphics[width=0.12\linewidth]{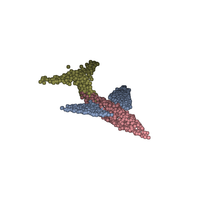}
\includegraphics[width=0.12\linewidth]{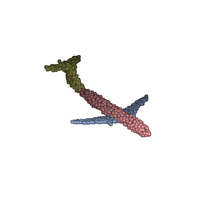}
\includegraphics[width=0.12\linewidth]{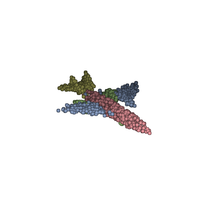}
\includegraphics[width=0.12\linewidth]{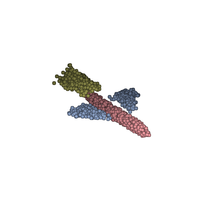}
\includegraphics[width=0.12\linewidth]{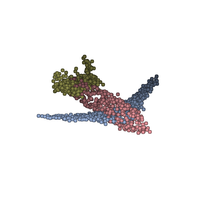}
\end{minipage}

\begin{minipage}{\linewidth}\centering
\includegraphics[width=0.12\linewidth]{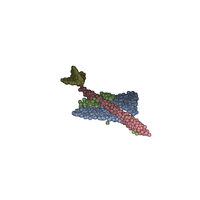}
\includegraphics[width=0.12\linewidth]{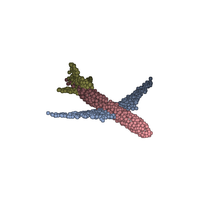}
\includegraphics[width=0.12\linewidth]{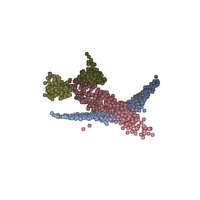}
\includegraphics[width=0.12\linewidth]{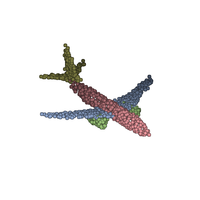}
\includegraphics[width=0.12\linewidth]{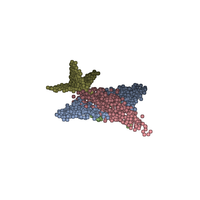}
\includegraphics[width=0.12\linewidth]{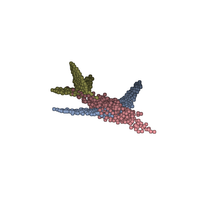}
\includegraphics[width=0.12\linewidth]{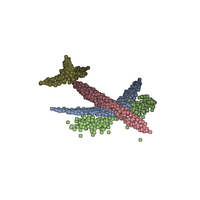}
\includegraphics[width=0.12\linewidth]{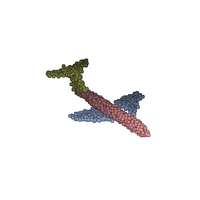}
\end{minipage}

\begin{minipage}{\linewidth}\centering
\includegraphics[width=0.12\linewidth]{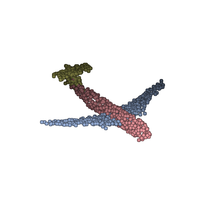}
\includegraphics[width=0.12\linewidth]{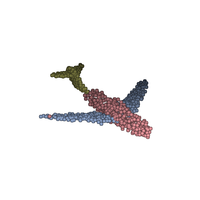}
\includegraphics[width=0.12\linewidth]{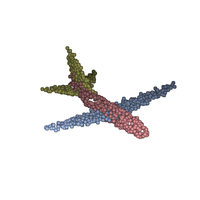}
\includegraphics[width=0.12\linewidth]{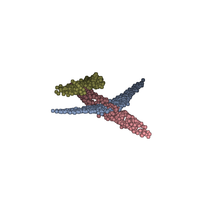}
\includegraphics[width=0.12\linewidth]{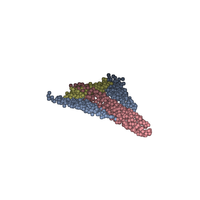}
\includegraphics[width=0.12\linewidth]{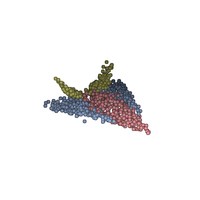}
\includegraphics[width=0.12\linewidth]{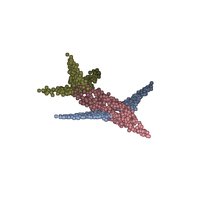}
\includegraphics[width=0.12\linewidth]{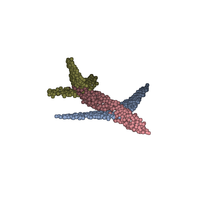}
\end{minipage}

\begin{minipage}{\linewidth}\centering
\includegraphics[width=0.12\linewidth]{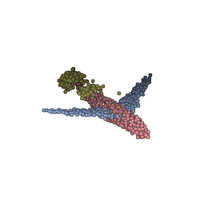}
\includegraphics[width=0.12\linewidth]{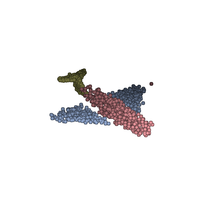}
\includegraphics[width=0.12\linewidth]{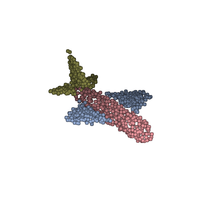}
\includegraphics[width=0.12\linewidth]{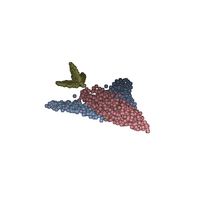}
\includegraphics[width=0.12\linewidth]{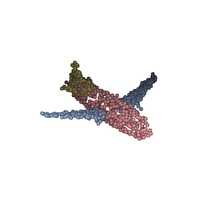}
\includegraphics[width=0.12\linewidth]{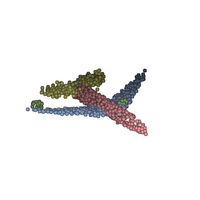}
\includegraphics[width=0.12\linewidth]{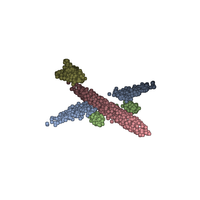}
\includegraphics[width=0.12\linewidth]{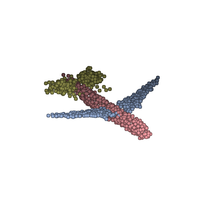}
\end{minipage}

	\caption{Airplane gallery. We present $80$ randomly sampled airplanes from the $5,000$ which were generated by \OurNet{}.}
	\label{fig:supp_plane_gallery}
\end{figure*}

%% file: vase_gallery.tex
\begin{figure*}\centering\begin{minipage}{\linewidth}\centering
\begin{minipage}{0.08\linewidth}\centering \includegraphics[width=0.5\linewidth]{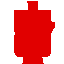} \end{minipage}\begin{minipage}{0.08\linewidth}\centering \includegraphics[width=0.5\linewidth]{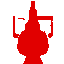} \end{minipage}\begin{minipage}{0.08\linewidth}\centering \includegraphics[width=0.5\linewidth]{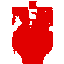} \end{minipage}\begin{minipage}{0.08\linewidth}\centering \includegraphics[width=0.5\linewidth]{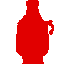} \end{minipage}\begin{minipage}{0.08\linewidth}\centering \includegraphics[width=0.5\linewidth]{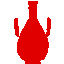} \end{minipage}\begin{minipage}{0.08\linewidth}\centering \includegraphics[width=0.5\linewidth]{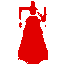} \end{minipage}\begin{minipage}{0.08\linewidth}\centering \includegraphics[width=0.5\linewidth]{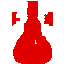} \end{minipage}\begin{minipage}{0.08\linewidth}\centering \includegraphics[width=0.5\linewidth]{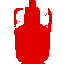} \end{minipage}\begin{minipage}{0.08\linewidth}\centering \includegraphics[width=0.5\linewidth]{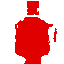} \end{minipage}\begin{minipage}{0.08\linewidth}\centering \includegraphics[width=0.5\linewidth]{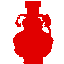} \end{minipage}\begin{minipage}{0.08\linewidth}\centering \includegraphics[width=0.5\linewidth]{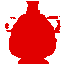} \end{minipage}\begin{minipage}{0.08\linewidth}\centering \includegraphics[width=0.5\linewidth]{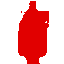} \end{minipage}\end{minipage}

\begin{minipage}{\linewidth}\centering
\begin{minipage}{0.08\linewidth}\centering \includegraphics[width=0.5\linewidth]{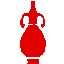} \end{minipage}\begin{minipage}{0.08\linewidth}\centering \includegraphics[width=0.5\linewidth]{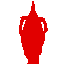} \end{minipage}\begin{minipage}{0.08\linewidth}\centering \includegraphics[width=0.5\linewidth]{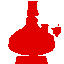} \end{minipage}\begin{minipage}{0.08\linewidth}\centering \includegraphics[width=0.5\linewidth]{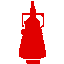} \end{minipage}\begin{minipage}{0.08\linewidth}\centering \includegraphics[width=0.5\linewidth]{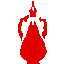} \end{minipage}\begin{minipage}{0.08\linewidth}\centering \includegraphics[width=0.5\linewidth]{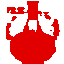} \end{minipage}\begin{minipage}{0.08\linewidth}\centering \includegraphics[width=0.5\linewidth]{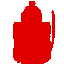} \end{minipage}\begin{minipage}{0.08\linewidth}\centering \includegraphics[width=0.5\linewidth]{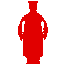} \end{minipage}\begin{minipage}{0.08\linewidth}\centering \includegraphics[width=0.5\linewidth]{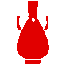} \end{minipage}\begin{minipage}{0.08\linewidth}\centering \includegraphics[width=0.5\linewidth]{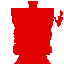} \end{minipage}\begin{minipage}{0.08\linewidth}\centering \includegraphics[width=0.5\linewidth]{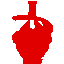} \end{minipage}\begin{minipage}{0.08\linewidth}\centering \includegraphics[width=0.5\linewidth]{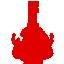} \end{minipage}\end{minipage}

\begin{minipage}{\linewidth}\centering
\begin{minipage}{0.08\linewidth}\centering \includegraphics[width=0.5\linewidth]{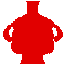} \end{minipage}\begin{minipage}{0.08\linewidth}\centering \includegraphics[width=0.5\linewidth]{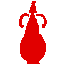} \end{minipage}\begin{minipage}{0.08\linewidth}\centering \includegraphics[width=0.5\linewidth]{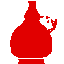} \end{minipage}\begin{minipage}{0.08\linewidth}\centering \includegraphics[width=0.5\linewidth]{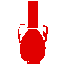} \end{minipage}\begin{minipage}{0.08\linewidth}\centering \includegraphics[width=0.5\linewidth]{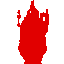} \end{minipage}\begin{minipage}{0.08\linewidth}\centering \includegraphics[width=0.5\linewidth]{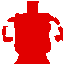} \end{minipage}\begin{minipage}{0.08\linewidth}\centering \includegraphics[width=0.5\linewidth]{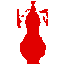} \end{minipage}\begin{minipage}{0.08\linewidth}\centering \includegraphics[width=0.5\linewidth]{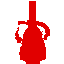} \end{minipage}\begin{minipage}{0.08\linewidth}\centering \includegraphics[width=0.5\linewidth]{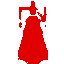} \end{minipage}\begin{minipage}{0.08\linewidth}\centering \includegraphics[width=0.5\linewidth]{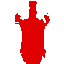} \end{minipage}\begin{minipage}{0.08\linewidth}\centering \includegraphics[width=0.5\linewidth]{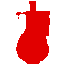} \end{minipage}\begin{minipage}{0.08\linewidth}\centering \includegraphics[width=0.5\linewidth]{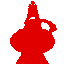} \end{minipage}\end{minipage}

\begin{minipage}{\linewidth}\centering
\begin{minipage}{0.08\linewidth}\centering \includegraphics[width=0.5\linewidth]{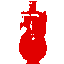} \end{minipage}\begin{minipage}{0.08\linewidth}\centering \includegraphics[width=0.5\linewidth]{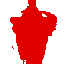} \end{minipage}\begin{minipage}{0.08\linewidth}\centering \includegraphics[width=0.5\linewidth]{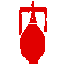} \end{minipage}\begin{minipage}{0.08\linewidth}\centering \includegraphics[width=0.5\linewidth]{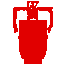} \end{minipage}\begin{minipage}{0.08\linewidth}\centering \includegraphics[width=0.5\linewidth]{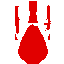} \end{minipage}\begin{minipage}{0.08\linewidth}\centering \includegraphics[width=0.5\linewidth]{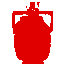} \end{minipage}\begin{minipage}{0.08\linewidth}\centering \includegraphics[width=0.5\linewidth]{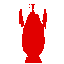} \end{minipage}\begin{minipage}{0.08\linewidth}\centering \includegraphics[width=0.5\linewidth]{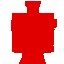} \end{minipage}\begin{minipage}{0.08\linewidth}\centering \includegraphics[width=0.5\linewidth]{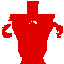} \end{minipage}\begin{minipage}{0.08\linewidth}\centering \includegraphics[width=0.5\linewidth]{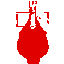} \end{minipage}\begin{minipage}{0.08\linewidth}\centering \includegraphics[width=0.5\linewidth]{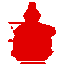} \end{minipage}\begin{minipage}{0.08\linewidth}\centering \includegraphics[width=0.5\linewidth]{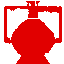} \end{minipage}\end{minipage}

\begin{minipage}{\linewidth}\centering
\begin{minipage}{0.08\linewidth}\centering \includegraphics[width=0.5\linewidth]{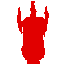} \end{minipage}\begin{minipage}{0.08\linewidth}\centering \includegraphics[width=0.5\linewidth]{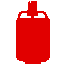} \end{minipage}\begin{minipage}{0.08\linewidth}\centering \includegraphics[width=0.5\linewidth]{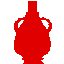} \end{minipage}\begin{minipage}{0.08\linewidth}\centering \includegraphics[width=0.5\linewidth]{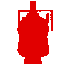} \end{minipage}\begin{minipage}{0.08\linewidth}\centering \includegraphics[width=0.5\linewidth]{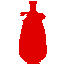} \end{minipage}\begin{minipage}{0.08\linewidth}\centering \includegraphics[width=0.5\linewidth]{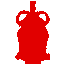} \end{minipage}\begin{minipage}{0.08\linewidth}\centering \includegraphics[width=0.5\linewidth]{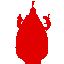} \end{minipage}\begin{minipage}{0.08\linewidth}\centering \includegraphics[width=0.5\linewidth]{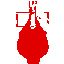} \end{minipage}\begin{minipage}{0.08\linewidth}\centering \includegraphics[width=0.5\linewidth]{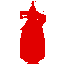} \end{minipage}\begin{minipage}{0.08\linewidth}\centering \includegraphics[width=0.5\linewidth]{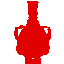} \end{minipage}\begin{minipage}{0.08\linewidth}\centering \includegraphics[width=0.5\linewidth]{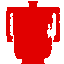} \end{minipage}\begin{minipage}{0.08\linewidth}\centering \includegraphics[width=0.5\linewidth]{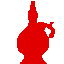} \end{minipage}\end{minipage}

\begin{minipage}{\linewidth}\centering
\begin{minipage}{0.08\linewidth}\centering \includegraphics[width=0.5\linewidth]{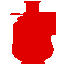} \end{minipage}\begin{minipage}{0.08\linewidth}\centering \includegraphics[width=0.5\linewidth]{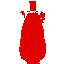} \end{minipage}\begin{minipage}{0.08\linewidth}\centering \includegraphics[width=0.5\linewidth]{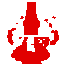} \end{minipage}\begin{minipage}{0.08\linewidth}\centering \includegraphics[width=0.5\linewidth]{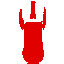} \end{minipage}\begin{minipage}{0.08\linewidth}\centering \includegraphics[width=0.5\linewidth]{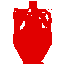} \end{minipage}\begin{minipage}{0.08\linewidth}\centering \includegraphics[width=0.5\linewidth]{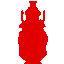} \end{minipage}\begin{minipage}{0.08\linewidth}\centering \includegraphics[width=0.5\linewidth]{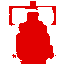} \end{minipage}\begin{minipage}{0.08\linewidth}\centering \includegraphics[width=0.5\linewidth]{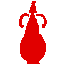} \end{minipage}\begin{minipage}{0.08\linewidth}\centering \includegraphics[width=0.5\linewidth]{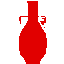} \end{minipage}\begin{minipage}{0.08\linewidth}\centering \includegraphics[width=0.5\linewidth]{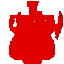} \end{minipage}\begin{minipage}{0.08\linewidth}\centering \includegraphics[width=0.5\linewidth]{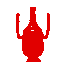} \end{minipage}\begin{minipage}{0.08\linewidth}\centering \includegraphics[width=0.5\linewidth]{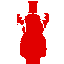} \end{minipage}\end{minipage}

\begin{minipage}{\linewidth}\centering
\begin{minipage}{0.08\linewidth}\centering \includegraphics[width=0.5\linewidth]{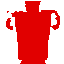} \end{minipage}\begin{minipage}{0.08\linewidth}\centering \includegraphics[width=0.5\linewidth]{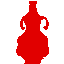} \end{minipage}\begin{minipage}{0.08\linewidth}\centering \includegraphics[width=0.5\linewidth]{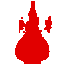} \end{minipage}\begin{minipage}{0.08\linewidth}\centering \includegraphics[width=0.5\linewidth]{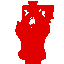} \end{minipage}\begin{minipage}{0.08\linewidth}\centering \includegraphics[width=0.5\linewidth]{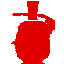} \end{minipage}\begin{minipage}{0.08\linewidth}\centering \includegraphics[width=0.5\linewidth]{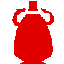} \end{minipage}\begin{minipage}{0.08\linewidth}\centering \includegraphics[width=0.5\linewidth]{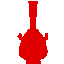} \end{minipage}\begin{minipage}{0.08\linewidth}\centering \includegraphics[width=0.5\linewidth]{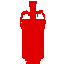} \end{minipage}\begin{minipage}{0.08\linewidth}\centering \includegraphics[width=0.5\linewidth]{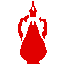} \end{minipage}\begin{minipage}{0.08\linewidth}\centering \includegraphics[width=0.5\linewidth]{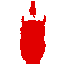} \end{minipage}\begin{minipage}{0.08\linewidth}\centering \includegraphics[width=0.5\linewidth]{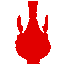} \end{minipage}\begin{minipage}{0.08\linewidth}\centering \includegraphics[width=0.5\linewidth]{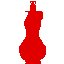} \end{minipage}\end{minipage}

\begin{minipage}{\linewidth}\centering
\begin{minipage}{0.08\linewidth}\centering \includegraphics[width=0.5\linewidth]{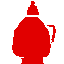} \end{minipage}\begin{minipage}{0.08\linewidth}\centering \includegraphics[width=0.5\linewidth]{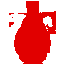} \end{minipage}\begin{minipage}{0.08\linewidth}\centering \includegraphics[width=0.5\linewidth]{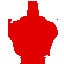} \end{minipage}\begin{minipage}{0.08\linewidth}\centering \includegraphics[width=0.5\linewidth]{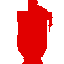} \end{minipage}\begin{minipage}{0.08\linewidth}\centering \includegraphics[width=0.5\linewidth]{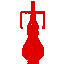} \end{minipage}\begin{minipage}{0.08\linewidth}\centering \includegraphics[width=0.5\linewidth]{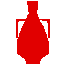} \end{minipage}\begin{minipage}{0.08\linewidth}\centering \includegraphics[width=0.5\linewidth]{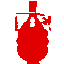} \end{minipage}\begin{minipage}{0.08\linewidth}\centering \includegraphics[width=0.5\linewidth]{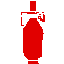} \end{minipage}\begin{minipage}{0.08\linewidth}\centering \includegraphics[width=0.5\linewidth]{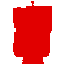} \end{minipage}\begin{minipage}{0.08\linewidth}\centering \includegraphics[width=0.5\linewidth]{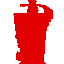} \end{minipage}\begin{minipage}{0.08\linewidth}\centering \includegraphics[width=0.5\linewidth]{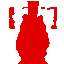} \end{minipage}\begin{minipage}{0.08\linewidth}\centering \includegraphics[width=0.5\linewidth]{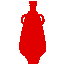} \end{minipage}\end{minipage}

\begin{minipage}{\linewidth}\centering
\begin{minipage}{0.08\linewidth}\centering \includegraphics[width=0.5\linewidth]{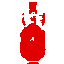} \end{minipage}\begin{minipage}{0.08\linewidth}\centering \includegraphics[width=0.5\linewidth]{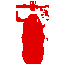} \end{minipage}\begin{minipage}{0.08\linewidth}\centering \includegraphics[width=0.5\linewidth]{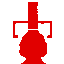} \end{minipage}\begin{minipage}{0.08\linewidth}\centering \includegraphics[width=0.5\linewidth]{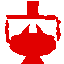} \end{minipage}\begin{minipage}{0.08\linewidth}\centering \includegraphics[width=0.5\linewidth]{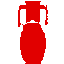} \end{minipage}\begin{minipage}{0.08\linewidth}\centering \includegraphics[width=0.5\linewidth]{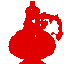} \end{minipage}\begin{minipage}{0.08\linewidth}\centering \includegraphics[width=0.5\linewidth]{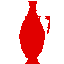} \end{minipage}\begin{minipage}{0.08\linewidth}\centering \includegraphics[width=0.5\linewidth]{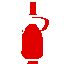} \end{minipage}\begin{minipage}{0.08\linewidth}\centering \includegraphics[width=0.5\linewidth]{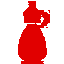} \end{minipage}\begin{minipage}{0.08\linewidth}\centering \includegraphics[width=0.5\linewidth]{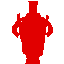} \end{minipage}\begin{minipage}{0.08\linewidth}\centering \includegraphics[width=0.5\linewidth]{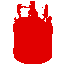} \end{minipage}\begin{minipage}{0.08\linewidth}\centering \includegraphics[width=0.5\linewidth]{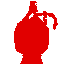} \end{minipage}\end{minipage}

\begin{minipage}{\linewidth}\centering
\begin{minipage}{0.08\linewidth}\centering \includegraphics[width=0.5\linewidth]{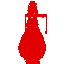} \end{minipage}\begin{minipage}{0.08\linewidth}\centering \includegraphics[width=0.5\linewidth]{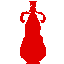} \end{minipage}\begin{minipage}{0.08\linewidth}\centering \includegraphics[width=0.5\linewidth]{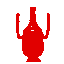} \end{minipage}\begin{minipage}{0.08\linewidth}\centering \includegraphics[width=0.5\linewidth]{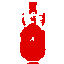} \end{minipage}\begin{minipage}{0.08\linewidth}\centering \includegraphics[width=0.5\linewidth]{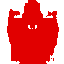} \end{minipage}\begin{minipage}{0.08\linewidth}\centering \includegraphics[width=0.5\linewidth]{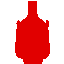} \end{minipage}\begin{minipage}{0.08\linewidth}\centering \includegraphics[width=0.5\linewidth]{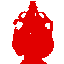} \end{minipage}\begin{minipage}{0.08\linewidth}\centering \includegraphics[width=0.5\linewidth]{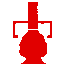} \end{minipage}\begin{minipage}{0.08\linewidth}\centering \includegraphics[width=0.5\linewidth]{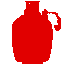} \end{minipage}\begin{minipage}{0.08\linewidth}\centering \includegraphics[width=0.5\linewidth]{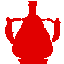} \end{minipage}\begin{minipage}{0.08\linewidth}\centering \includegraphics[width=0.5\linewidth]{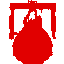} \end{minipage}\begin{minipage}{0.08\linewidth}\centering \includegraphics[width=0.5\linewidth]{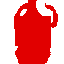} \end{minipage}\end{minipage}

\begin{minipage}{\linewidth}\centering
\begin{minipage}{0.08\linewidth}\centering \includegraphics[width=0.5\linewidth]{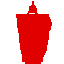} \end{minipage}\begin{minipage}{0.08\linewidth}\centering \includegraphics[width=0.5\linewidth]{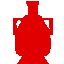} \end{minipage}\begin{minipage}{0.08\linewidth}\centering \includegraphics[width=0.5\linewidth]{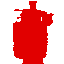} \end{minipage}\begin{minipage}{0.08\linewidth}\centering \includegraphics[width=0.5\linewidth]{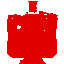} \end{minipage}\begin{minipage}{0.08\linewidth}\centering \includegraphics[width=0.5\linewidth]{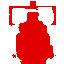} \end{minipage}\begin{minipage}{0.08\linewidth}\centering \includegraphics[width=0.5\linewidth]{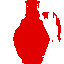} \end{minipage}\begin{minipage}{0.08\linewidth}\centering \includegraphics[width=0.5\linewidth]{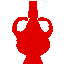} \end{minipage}\begin{minipage}{0.08\linewidth}\centering \includegraphics[width=0.5\linewidth]{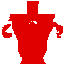} \end{minipage}\begin{minipage}{0.08\linewidth}\centering \includegraphics[width=0.5\linewidth]{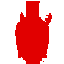} \end{minipage}\begin{minipage}{0.08\linewidth}\centering \includegraphics[width=0.5\linewidth]{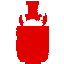} \end{minipage}\begin{minipage}{0.08\linewidth}\centering \includegraphics[width=0.5\linewidth]{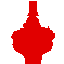} \end{minipage}\begin{minipage}{0.08\linewidth}\centering \includegraphics[width=0.5\linewidth]{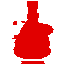} \end{minipage}\end{minipage}

\begin{minipage}{\linewidth}\centering
\begin{minipage}{0.08\linewidth}\centering \includegraphics[width=0.5\linewidth]{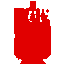} \end{minipage}\begin{minipage}{0.08\linewidth}\centering \includegraphics[width=0.5\linewidth]{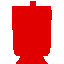} \end{minipage}\begin{minipage}{0.08\linewidth}\centering \includegraphics[width=0.5\linewidth]{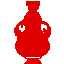} \end{minipage}\begin{minipage}{0.08\linewidth}\centering \includegraphics[width=0.5\linewidth]{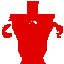} \end{minipage}\begin{minipage}{0.08\linewidth}\centering \includegraphics[width=0.5\linewidth]{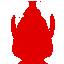} \end{minipage}\begin{minipage}{0.08\linewidth}\centering \includegraphics[width=0.5\linewidth]{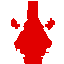} \end{minipage}\begin{minipage}{0.08\linewidth}\centering \includegraphics[width=0.5\linewidth]{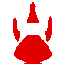} \end{minipage}\begin{minipage}{0.08\linewidth}\centering \includegraphics[width=0.5\linewidth]{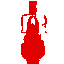} \end{minipage}\begin{minipage}{0.08\linewidth}\centering \includegraphics[width=0.5\linewidth]{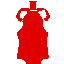} \end{minipage}\begin{minipage}{0.08\linewidth}\centering \includegraphics[width=0.5\linewidth]{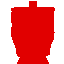} \end{minipage}\begin{minipage}{0.08\linewidth}\centering \includegraphics[width=0.5\linewidth]{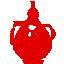} \end{minipage}\begin{minipage}{0.08\linewidth}\centering \includegraphics[width=0.5\linewidth]{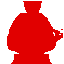} \end{minipage}\end{minipage}

\begin{minipage}{\linewidth}\centering
\begin{minipage}{0.08\linewidth}\centering \includegraphics[width=0.5\linewidth]{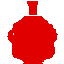} \end{minipage}\begin{minipage}{0.08\linewidth}\centering \includegraphics[width=0.5\linewidth]{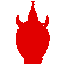} \end{minipage}\begin{minipage}{0.08\linewidth}\centering \includegraphics[width=0.5\linewidth]{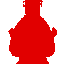} \end{minipage}\begin{minipage}{0.08\linewidth}\centering \includegraphics[width=0.5\linewidth]{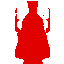} \end{minipage}\begin{minipage}{0.08\linewidth}\centering \includegraphics[width=0.5\linewidth]{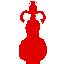} \end{minipage}\begin{minipage}{0.08\linewidth}\centering \includegraphics[width=0.5\linewidth]{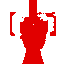} \end{minipage}\begin{minipage}{0.08\linewidth}\centering \includegraphics[width=0.5\linewidth]{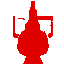} \end{minipage}\begin{minipage}{0.08\linewidth}\centering \includegraphics[width=0.5\linewidth]{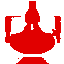} \end{minipage}\begin{minipage}{0.08\linewidth}\centering \includegraphics[width=0.5\linewidth]{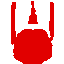} \end{minipage}\begin{minipage}{0.08\linewidth}\centering \includegraphics[width=0.5\linewidth]{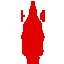} \end{minipage}\begin{minipage}{0.08\linewidth}\centering \includegraphics[width=0.5\linewidth]{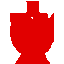} \end{minipage}\begin{minipage}{0.08\linewidth}\centering \includegraphics[width=0.5\linewidth]{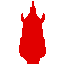} \end{minipage}\end{minipage}

\begin{minipage}{\linewidth}\centering
\begin{minipage}{0.08\linewidth}\centering \includegraphics[width=0.5\linewidth]{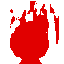} \end{minipage}\begin{minipage}{0.08\linewidth}\centering \includegraphics[width=0.5\linewidth]{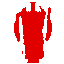} \end{minipage}\begin{minipage}{0.08\linewidth}\centering \includegraphics[width=0.5\linewidth]{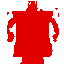} \end{minipage}\begin{minipage}{0.08\linewidth}\centering \includegraphics[width=0.5\linewidth]{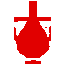} \end{minipage}\begin{minipage}{0.08\linewidth}\centering \includegraphics[width=0.5\linewidth]{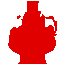} \end{minipage}\begin{minipage}{0.08\linewidth}\centering \includegraphics[width=0.5\linewidth]{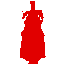} \end{minipage}\begin{minipage}{0.08\linewidth}\centering \includegraphics[width=0.5\linewidth]{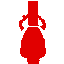} \end{minipage}\begin{minipage}{0.08\linewidth}\centering \includegraphics[width=0.5\linewidth]{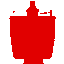} \end{minipage}\begin{minipage}{0.08\linewidth}\centering \includegraphics[width=0.5\linewidth]{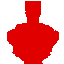} \end{minipage}\begin{minipage}{0.08\linewidth}\centering \includegraphics[width=0.5\linewidth]{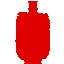} \end{minipage}\begin{minipage}{0.08\linewidth}\centering \includegraphics[width=0.5\linewidth]{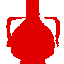} \end{minipage}\begin{minipage}{0.08\linewidth}\centering \includegraphics[width=0.5\linewidth]{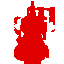} \end{minipage}\end{minipage}

\begin{minipage}{\linewidth}\centering
\begin{minipage}{0.08\linewidth}\centering \includegraphics[width=0.5\linewidth]{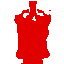} \end{minipage}\begin{minipage}{0.08\linewidth}\centering \includegraphics[width=0.5\linewidth]{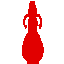} \end{minipage}\begin{minipage}{0.08\linewidth}\centering \includegraphics[width=0.5\linewidth]{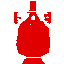} \end{minipage}\begin{minipage}{0.08\linewidth}\centering \includegraphics[width=0.5\linewidth]{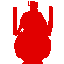} \end{minipage}\begin{minipage}{0.08\linewidth}\centering \includegraphics[width=0.5\linewidth]{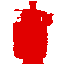} \end{minipage}\begin{minipage}{0.08\linewidth}\centering \includegraphics[width=0.5\linewidth]{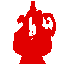} \end{minipage}\begin{minipage}{0.08\linewidth}\centering \includegraphics[width=0.5\linewidth]{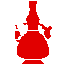} \end{minipage}\begin{minipage}{0.08\linewidth}\centering \includegraphics[width=0.5\linewidth]{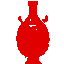} \end{minipage}\begin{minipage}{0.08\linewidth}\centering \includegraphics[width=0.5\linewidth]{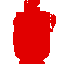} \end{minipage}\begin{minipage}{0.08\linewidth}\centering \includegraphics[width=0.5\linewidth]{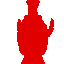} \end{minipage}\begin{minipage}{0.08\linewidth}\centering \includegraphics[width=0.5\linewidth]{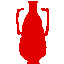} \end{minipage}\begin{minipage}{0.08\linewidth}\centering \includegraphics[width=0.5\linewidth]{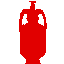} \end{minipage}\end{minipage}

\begin{minipage}{\linewidth}\centering
\begin{minipage}{0.08\linewidth}\centering \includegraphics[width=0.5\linewidth]{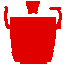} \end{minipage}\begin{minipage}{0.08\linewidth}\centering \includegraphics[width=0.5\linewidth]{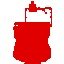} \end{minipage}\begin{minipage}{0.08\linewidth}\centering \includegraphics[width=0.5\linewidth]{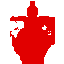} \end{minipage}\begin{minipage}{0.08\linewidth}\centering \includegraphics[width=0.5\linewidth]{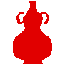} \end{minipage}\begin{minipage}{0.08\linewidth}\centering \includegraphics[width=0.5\linewidth]{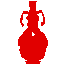} \end{minipage}\begin{minipage}{0.08\linewidth}\centering \includegraphics[width=0.5\linewidth]{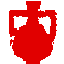} \end{minipage}\begin{minipage}{0.08\linewidth}\centering \includegraphics[width=0.5\linewidth]{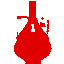} \end{minipage}\begin{minipage}{0.08\linewidth}\centering \includegraphics[width=0.5\linewidth]{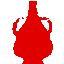} \end{minipage}\begin{minipage}{0.08\linewidth}\centering \includegraphics[width=0.5\linewidth]{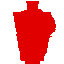} \end{minipage}\begin{minipage}{0.08\linewidth}\centering \includegraphics[width=0.5\linewidth]{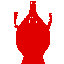} \end{minipage}\begin{minipage}{0.08\linewidth}\centering \includegraphics[width=0.5\linewidth]{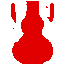} \end{minipage}\begin{minipage}{0.08\linewidth}\centering \includegraphics[width=0.5\linewidth]{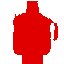} \end{minipage}\end{minipage}

\begin{minipage}{\linewidth}\centering
\begin{minipage}{0.08\linewidth}\centering \includegraphics[width=0.5\linewidth]{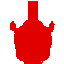} \end{minipage}\begin{minipage}{0.08\linewidth}\centering \includegraphics[width=0.5\linewidth]{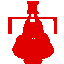} \end{minipage}\begin{minipage}{0.08\linewidth}\centering \includegraphics[width=0.5\linewidth]{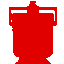} \end{minipage}\begin{minipage}{0.08\linewidth}\centering \includegraphics[width=0.5\linewidth]{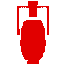} \end{minipage}\begin{minipage}{0.08\linewidth}\centering \includegraphics[width=0.5\linewidth]{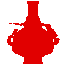} \end{minipage}\begin{minipage}{0.08\linewidth}\centering \includegraphics[width=0.5\linewidth]{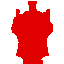} \end{minipage}\begin{minipage}{0.08\linewidth}\centering \includegraphics[width=0.5\linewidth]{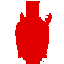} \end{minipage}\begin{minipage}{0.08\linewidth}\centering \includegraphics[width=0.5\linewidth]{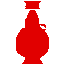} \end{minipage}\begin{minipage}{0.08\linewidth}\centering \includegraphics[width=0.5\linewidth]{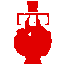} \end{minipage}\begin{minipage}{0.08\linewidth}\centering \includegraphics[width=0.5\linewidth]{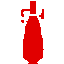} \end{minipage}\begin{minipage}{0.08\linewidth}\centering \includegraphics[width=0.5\linewidth]{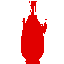} \end{minipage}\begin{minipage}{0.08\linewidth}\centering \includegraphics[width=0.5\linewidth]{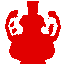} \end{minipage}\end{minipage}

\begin{minipage}{\linewidth}\centering
\begin{minipage}{0.08\linewidth}\centering \includegraphics[width=0.5\linewidth]{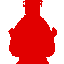} \end{minipage}\begin{minipage}{0.08\linewidth}\centering \includegraphics[width=0.5\linewidth]{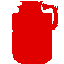} \end{minipage}\begin{minipage}{0.08\linewidth}\centering \includegraphics[width=0.5\linewidth]{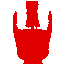} \end{minipage}\begin{minipage}{0.08\linewidth}\centering \includegraphics[width=0.5\linewidth]{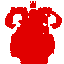} \end{minipage}\begin{minipage}{0.08\linewidth}\centering \includegraphics[width=0.5\linewidth]{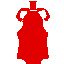} \end{minipage}\begin{minipage}{0.08\linewidth}\centering \includegraphics[width=0.5\linewidth]{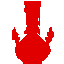} \end{minipage}\begin{minipage}{0.08\linewidth}\centering \includegraphics[width=0.5\linewidth]{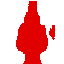} \end{minipage}\begin{minipage}{0.08\linewidth}\centering \includegraphics[width=0.5\linewidth]{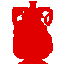} \end{minipage}\begin{minipage}{0.08\linewidth}\centering \includegraphics[width=0.5\linewidth]{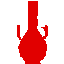} \end{minipage}\begin{minipage}{0.08\linewidth}\centering \includegraphics[width=0.5\linewidth]{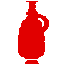} \end{minipage}\begin{minipage}{0.08\linewidth}\centering \includegraphics[width=0.5\linewidth]{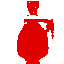} \end{minipage}\begin{minipage}{0.08\linewidth}\centering \includegraphics[width=0.5\linewidth]{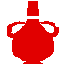} \end{minipage}\end{minipage}

\begin{minipage}{\linewidth}\centering
\begin{minipage}{0.08\linewidth}\centering \includegraphics[width=0.5\linewidth]{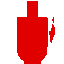} \end{minipage}\begin{minipage}{0.08\linewidth}\centering \includegraphics[width=0.5\linewidth]{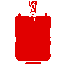} \end{minipage}\begin{minipage}{0.08\linewidth}\centering \includegraphics[width=0.5\linewidth]{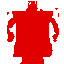} \end{minipage}\begin{minipage}{0.08\linewidth}\centering \includegraphics[width=0.5\linewidth]{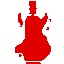} \end{minipage}\begin{minipage}{0.08\linewidth}\centering \includegraphics[width=0.5\linewidth]{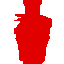} \end{minipage}\begin{minipage}{0.08\linewidth}\centering \includegraphics[width=0.5\linewidth]{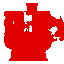} \end{minipage}\begin{minipage}{0.08\linewidth}\centering \includegraphics[width=0.5\linewidth]{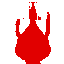} \end{minipage}\begin{minipage}{0.08\linewidth}\centering \includegraphics[width=0.5\linewidth]{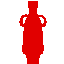} \end{minipage}\begin{minipage}{0.08\linewidth}\centering \includegraphics[width=0.5\linewidth]{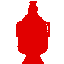} \end{minipage}\begin{minipage}{0.08\linewidth}\centering \includegraphics[width=0.5\linewidth]{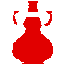} \end{minipage}\begin{minipage}{0.08\linewidth}\centering \includegraphics[width=0.5\linewidth]{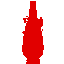} \end{minipage}\begin{minipage}{0.08\linewidth}\centering \includegraphics[width=0.5\linewidth]{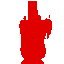} \end{minipage}\end{minipage}

\begin{minipage}{\linewidth}\centering
\begin{minipage}{0.08\linewidth}\centering \includegraphics[width=0.5\linewidth]{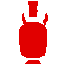} \end{minipage}\begin{minipage}{0.08\linewidth}\centering \includegraphics[width=0.5\linewidth]{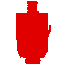} \end{minipage}\begin{minipage}{0.08\linewidth}\centering \includegraphics[width=0.5\linewidth]{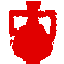} \end{minipage}\begin{minipage}{0.08\linewidth}\centering \includegraphics[width=0.5\linewidth]{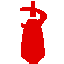} \end{minipage}\begin{minipage}{0.08\linewidth}\centering \includegraphics[width=0.5\linewidth]{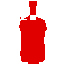} \end{minipage}\begin{minipage}{0.08\linewidth}\centering \includegraphics[width=0.5\linewidth]{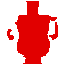} \end{minipage}\begin{minipage}{0.08\linewidth}\centering \includegraphics[width=0.5\linewidth]{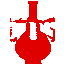} \end{minipage}\begin{minipage}{0.08\linewidth}\centering \includegraphics[width=0.5\linewidth]{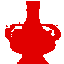} \end{minipage}\begin{minipage}{0.08\linewidth}\centering \includegraphics[width=0.5\linewidth]{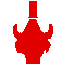} \end{minipage}\begin{minipage}{0.08\linewidth}\centering \includegraphics[width=0.5\linewidth]{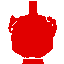} \end{minipage}\begin{minipage}{0.08\linewidth}\centering \includegraphics[width=0.5\linewidth]{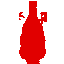} \end{minipage}\begin{minipage}{0.08\linewidth}\centering \includegraphics[width=0.5\linewidth]{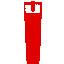} \end{minipage}\end{minipage}

\begin{minipage}{\linewidth}\centering
\begin{minipage}{0.08\linewidth}\centering \includegraphics[width=0.5\linewidth]{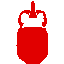} \end{minipage}\begin{minipage}{0.08\linewidth}\centering \includegraphics[width=0.5\linewidth]{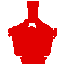} \end{minipage}\begin{minipage}{0.08\linewidth}\centering \includegraphics[width=0.5\linewidth]{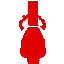} \end{minipage}\begin{minipage}{0.08\linewidth}\centering \includegraphics[width=0.5\linewidth]{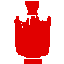} \end{minipage}\begin{minipage}{0.08\linewidth}\centering \includegraphics[width=0.5\linewidth]{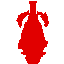} \end{minipage}\begin{minipage}{0.08\linewidth}\centering \includegraphics[width=0.5\linewidth]{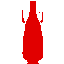} \end{minipage}\begin{minipage}{0.08\linewidth}\centering \includegraphics[width=0.5\linewidth]{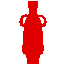} \end{minipage}\begin{minipage}{0.08\linewidth}\centering \includegraphics[width=0.5\linewidth]{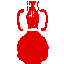} \end{minipage}\begin{minipage}{0.08\linewidth}\centering \includegraphics[width=0.5\linewidth]{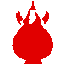} \end{minipage}\begin{minipage}{0.08\linewidth}\centering \includegraphics[width=0.5\linewidth]{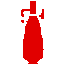} \end{minipage}\begin{minipage}{0.08\linewidth}\centering \includegraphics[width=0.5\linewidth]{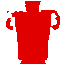} \end{minipage}\begin{minipage}{0.08\linewidth}\centering \includegraphics[width=0.5\linewidth]{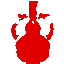} \end{minipage}\end{minipage}

\begin{minipage}{\linewidth}\centering
\begin{minipage}{0.08\linewidth}\centering \includegraphics[width=0.5\linewidth]{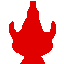} \end{minipage}\begin{minipage}{0.08\linewidth}\centering \includegraphics[width=0.5\linewidth]{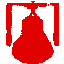} \end{minipage}\begin{minipage}{0.08\linewidth}\centering \includegraphics[width=0.5\linewidth]{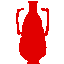} \end{minipage}\begin{minipage}{0.08\linewidth}\centering \includegraphics[width=0.5\linewidth]{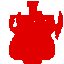} \end{minipage}\begin{minipage}{0.08\linewidth}\centering \includegraphics[width=0.5\linewidth]{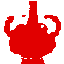} \end{minipage}\begin{minipage}{0.08\linewidth}\centering \includegraphics[width=0.5\linewidth]{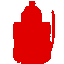} \end{minipage}\begin{minipage}{0.08\linewidth}\centering \includegraphics[width=0.5\linewidth]{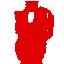} \end{minipage}\begin{minipage}{0.08\linewidth}\centering \includegraphics[width=0.5\linewidth]{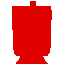} \end{minipage}\begin{minipage}{0.08\linewidth}\centering \includegraphics[width=0.5\linewidth]{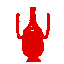} \end{minipage}\begin{minipage}{0.08\linewidth}\centering \includegraphics[width=0.5\linewidth]{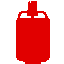} \end{minipage}\begin{minipage}{0.08\linewidth}\centering \includegraphics[width=0.5\linewidth]{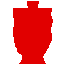} \end{minipage}\begin{minipage}{0.08\linewidth}\centering \includegraphics[width=0.5\linewidth]{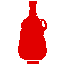} \end{minipage}\end{minipage}

\begin{minipage}{\linewidth}\centering
\begin{minipage}{0.08\linewidth}\centering \includegraphics[width=0.5\linewidth]{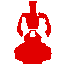} \end{minipage}\begin{minipage}{0.08\linewidth}\centering \includegraphics[width=0.5\linewidth]{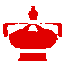} \end{minipage}\begin{minipage}{0.08\linewidth}\centering \includegraphics[width=0.5\linewidth]{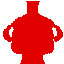} \end{minipage}\begin{minipage}{0.08\linewidth}\centering \includegraphics[width=0.5\linewidth]{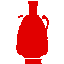} \end{minipage}\begin{minipage}{0.08\linewidth}\centering \includegraphics[width=0.5\linewidth]{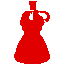} \end{minipage}\begin{minipage}{0.08\linewidth}\centering \includegraphics[width=0.5\linewidth]{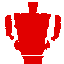} \end{minipage}\begin{minipage}{0.08\linewidth}\centering \includegraphics[width=0.5\linewidth]{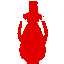} \end{minipage}\begin{minipage}{0.08\linewidth}\centering \includegraphics[width=0.5\linewidth]{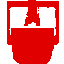} \end{minipage}\begin{minipage}{0.08\linewidth}\centering \includegraphics[width=0.5\linewidth]{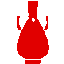} \end{minipage}\begin{minipage}{0.08\linewidth}\centering \includegraphics[width=0.5\linewidth]{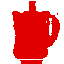} \end{minipage}\begin{minipage}{0.08\linewidth}\centering \includegraphics[width=0.5\linewidth]{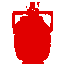} \end{minipage}\begin{minipage}{0.08\linewidth}\centering \includegraphics[width=0.5\linewidth]{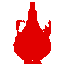} \end{minipage}\end{minipage}

\begin{minipage}{\linewidth}\centering
\begin{minipage}{0.08\linewidth}\centering \includegraphics[width=0.5\linewidth]{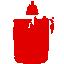} \end{minipage}\begin{minipage}{0.08\linewidth}\centering \includegraphics[width=0.5\linewidth]{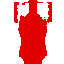} \end{minipage}\begin{minipage}{0.08\linewidth}\centering \includegraphics[width=0.5\linewidth]{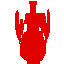} \end{minipage}\begin{minipage}{0.08\linewidth}\centering \includegraphics[width=0.5\linewidth]{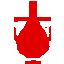} \end{minipage}\begin{minipage}{0.08\linewidth}\centering \includegraphics[width=0.5\linewidth]{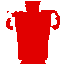} \end{minipage}\begin{minipage}{0.08\linewidth}\centering \includegraphics[width=0.5\linewidth]{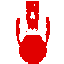} \end{minipage}\begin{minipage}{0.08\linewidth}\centering \includegraphics[width=0.5\linewidth]{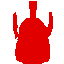} \end{minipage}\begin{minipage}{0.08\linewidth}\centering \includegraphics[width=0.5\linewidth]{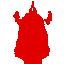} \end{minipage}\begin{minipage}{0.08\linewidth}\centering \includegraphics[width=0.5\linewidth]{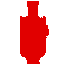} \end{minipage}\begin{minipage}{0.08\linewidth}\centering \includegraphics[width=0.5\linewidth]{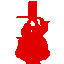} \end{minipage}\begin{minipage}{0.08\linewidth}\centering \includegraphics[width=0.5\linewidth]{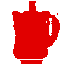} \end{minipage}\begin{minipage}{0.08\linewidth}\centering \includegraphics[width=0.5\linewidth]{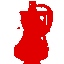} \end{minipage}\end{minipage}

\begin{minipage}{\linewidth}\centering
\begin{minipage}{0.08\linewidth}\centering \includegraphics[width=0.5\linewidth]{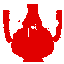} \end{minipage}\begin{minipage}{0.08\linewidth}\centering \includegraphics[width=0.5\linewidth]{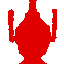} \end{minipage}\begin{minipage}{0.08\linewidth}\centering \includegraphics[width=0.5\linewidth]{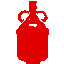} \end{minipage}\begin{minipage}{0.08\linewidth}\centering \includegraphics[width=0.5\linewidth]{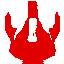} \end{minipage}\begin{minipage}{0.08\linewidth}\centering \includegraphics[width=0.5\linewidth]{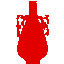} \end{minipage}\begin{minipage}{0.08\linewidth}\centering \includegraphics[width=0.5\linewidth]{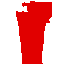} \end{minipage}\begin{minipage}{0.08\linewidth}\centering \includegraphics[width=0.5\linewidth]{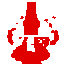} \end{minipage}\begin{minipage}{0.08\linewidth}\centering \includegraphics[width=0.5\linewidth]{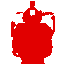} \end{minipage}\begin{minipage}{0.08\linewidth}\centering \includegraphics[width=0.5\linewidth]{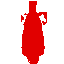} \end{minipage}\begin{minipage}{0.08\linewidth}\centering \includegraphics[width=0.5\linewidth]{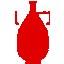} \end{minipage}\begin{minipage}{0.08\linewidth}\centering \includegraphics[width=0.5\linewidth]{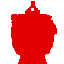} \end{minipage}\begin{minipage}{0.08\linewidth}\centering \includegraphics[width=0.5\linewidth]{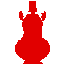} \end{minipage}\end{minipage}

	\caption{Vases gallery. We present $300$ randomly sampled vases from the $1,024$ which were generated by \OurNet{}.}
	\label{fig:supp_vase_gallery}
	\end{figure*}

%% file: supp_interpolation_chair.tex
\begin{figure*}\centering
\begin{minipage}{\linewidth}\centering
\includegraphics[width=0.19\linewidth]{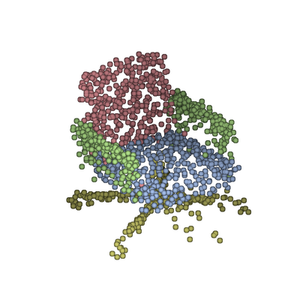}
\includegraphics[width=0.19\linewidth]{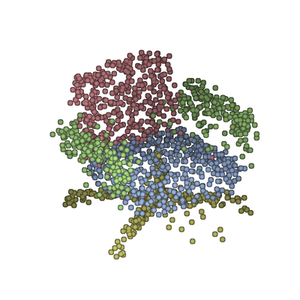}
\includegraphics[width=0.19\linewidth]{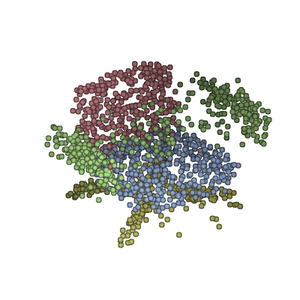}
\includegraphics[width=0.19\linewidth]{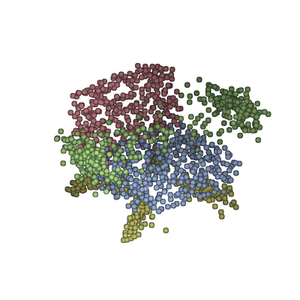}
\includegraphics[width=0.19\linewidth]{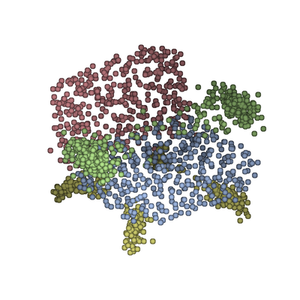}
\end{minipage}

\begin{minipage}{\linewidth}\centering
\includegraphics[width=0.19\linewidth]{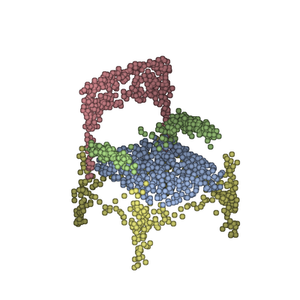}
\includegraphics[width=0.19\linewidth]{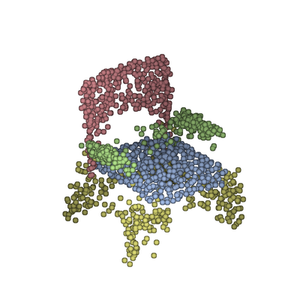}
\includegraphics[width=0.19\linewidth]{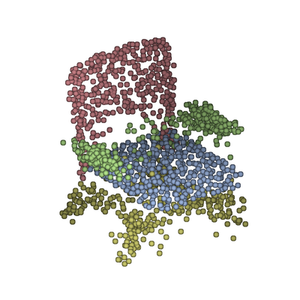}
\includegraphics[width=0.19\linewidth]{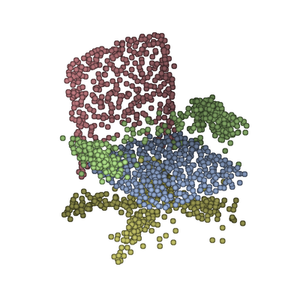}
\includegraphics[width=0.19\linewidth]{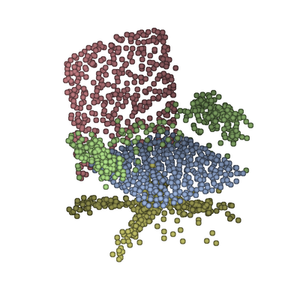}
\end{minipage}

\begin{minipage}{\linewidth}\centering
\includegraphics[width=0.19\linewidth]{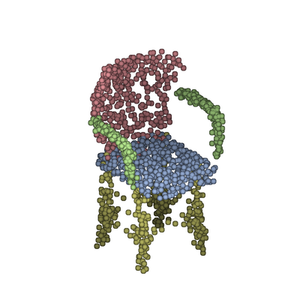}
\includegraphics[width=0.19\linewidth]{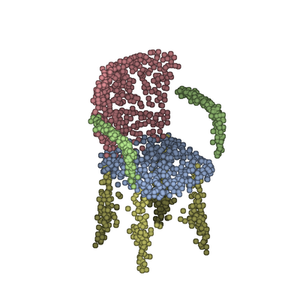}
\includegraphics[width=0.19\linewidth]{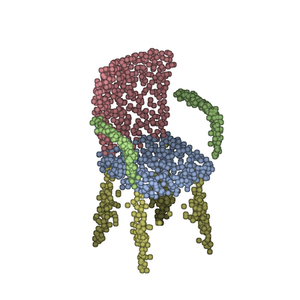}
\includegraphics[width=0.19\linewidth]{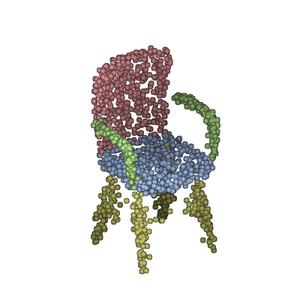}
\includegraphics[width=0.19\linewidth]{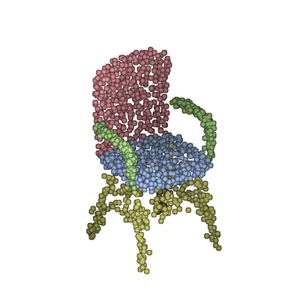}
\end{minipage}

\begin{minipage}{\linewidth}\centering
\includegraphics[width=0.19\linewidth]{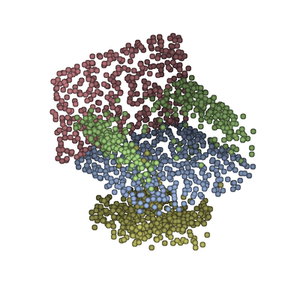}
\includegraphics[width=0.19\linewidth]{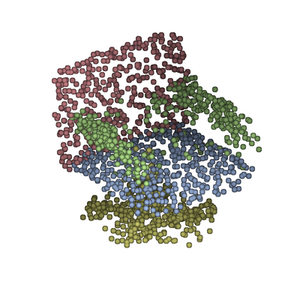}
\includegraphics[width=0.19\linewidth]{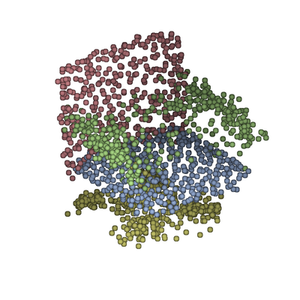}
\includegraphics[width=0.19\linewidth]{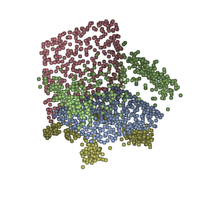}
\includegraphics[width=0.19\linewidth]{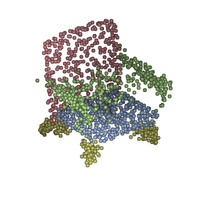}
\end{minipage}

\begin{minipage}{\linewidth}\centering
\includegraphics[width=0.19\linewidth]{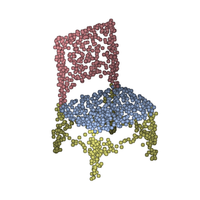}
\includegraphics[width=0.19\linewidth]{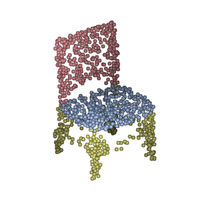}
\includegraphics[width=0.19\linewidth]{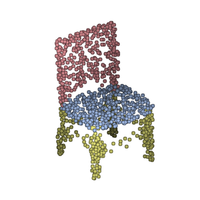}
\includegraphics[width=0.19\linewidth]{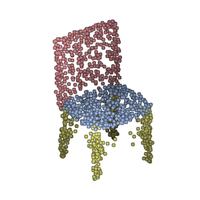}
\includegraphics[width=0.19\linewidth]{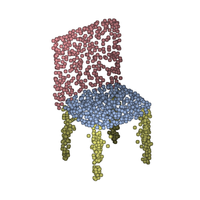}
\end{minipage}

\begin{minipage}{\linewidth}\centering
\includegraphics[width=0.19\linewidth]{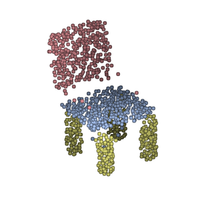}
\includegraphics[width=0.19\linewidth]{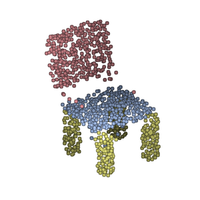}
\includegraphics[width=0.19\linewidth]{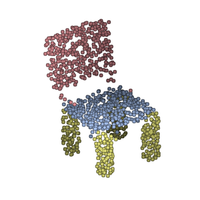}
\includegraphics[width=0.19\linewidth]{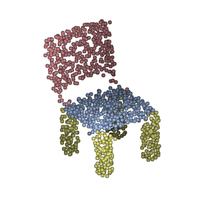}
\includegraphics[width=0.19\linewidth]{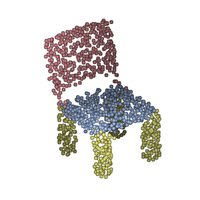}
\end{minipage}

	\caption{Complete latent space interpolation (parts and composition) for the chair category. The shape's code on the left is linearly interpolated to the shape's code on the right.}
	\label{fig:supp_interpolation_chair_full}
	\end{figure*}

\begin{figure*}\centering
\begin{minipage}{\linewidth}\centering
\includegraphics[width=0.19\linewidth]{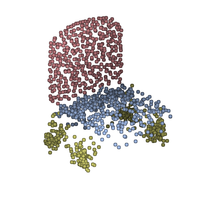}
\includegraphics[width=0.19\linewidth]{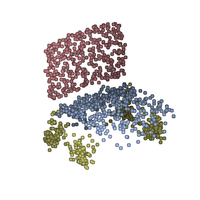}
\includegraphics[width=0.19\linewidth]{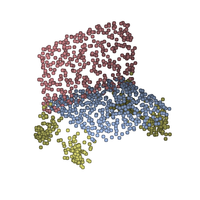}
\includegraphics[width=0.19\linewidth]{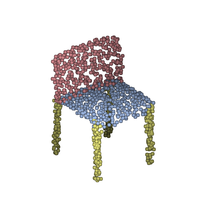}
\includegraphics[width=0.19\linewidth]{figures/chairs/part_interpolation/op_8_3.png}
\end{minipage}

\begin{minipage}{\linewidth}\centering
\includegraphics[width=0.19\linewidth]{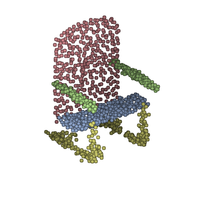}
\includegraphics[width=0.19\linewidth]{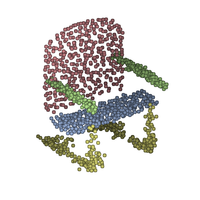}
\includegraphics[width=0.19\linewidth]{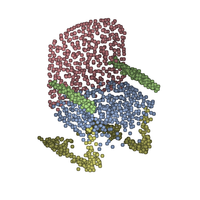}
\includegraphics[width=0.19\linewidth]{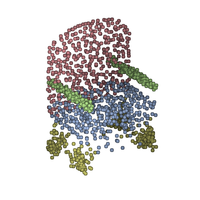}
\includegraphics[width=0.19\linewidth]{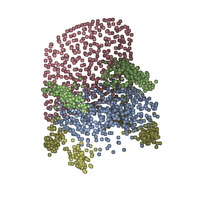}
\end{minipage}

\begin{minipage}{\linewidth}\centering
\includegraphics[width=0.19\linewidth]{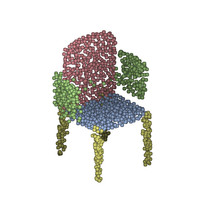}
\includegraphics[width=0.19\linewidth]{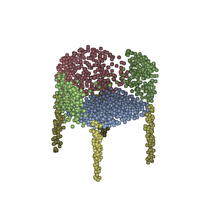}
\includegraphics[width=0.19\linewidth]{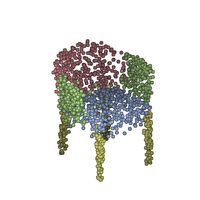}
\includegraphics[width=0.19\linewidth]{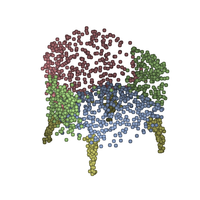}
\includegraphics[width=0.19\linewidth]{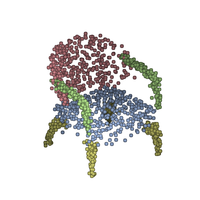}
\end{minipage}

\begin{minipage}{\linewidth}\centering
\includegraphics[width=0.19\linewidth]{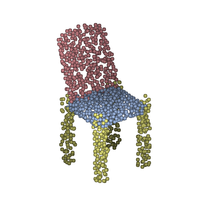}
\includegraphics[width=0.19\linewidth]{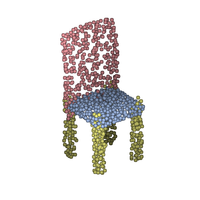}
\includegraphics[width=0.19\linewidth]{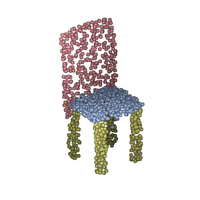}
\includegraphics[width=0.19\linewidth]{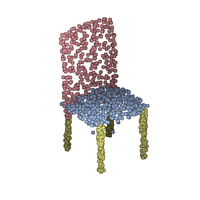}
\includegraphics[width=0.19\linewidth]{figures/chairs/part_interpolation/op_7_3.png}
\end{minipage}

	\caption{Chairs part-by-part interpolation. In each row we present part-by-part interpolation between two chairs (see left and right objects). The order of replacement is back-seat-legs-armrests. Please note, for certain pairs the arm-rests part does not exist, thus the interpolation of the arm-rest part is identity (no change).}
	\label{fig:supp_interpolation_chair_part}
	\end{figure*}

%% file: supp_interpolation_plane.tex
\begin{figure*}\centering
\begin{minipage}{\linewidth}\centering
\includegraphics[width=0.19\linewidth]{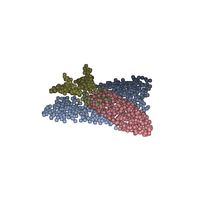}
\includegraphics[width=0.19\linewidth]{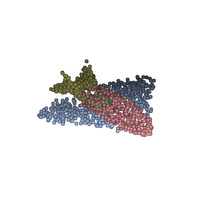}
\includegraphics[width=0.19\linewidth]{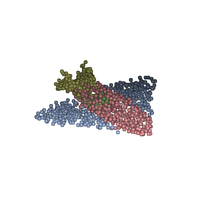}
\includegraphics[width=0.19\linewidth]{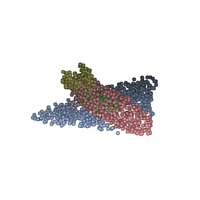}
\includegraphics[width=0.19\linewidth]{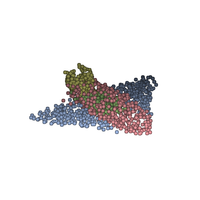}
\end{minipage}

\begin{minipage}{\linewidth}\centering
\includegraphics[width=0.19\linewidth]{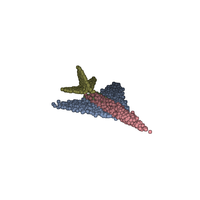}
\includegraphics[width=0.19\linewidth]{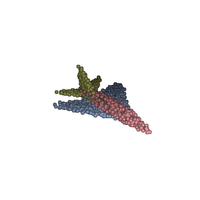}
\includegraphics[width=0.19\linewidth]{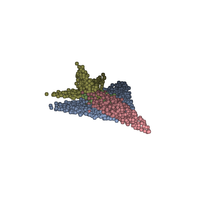}
\includegraphics[width=0.19\linewidth]{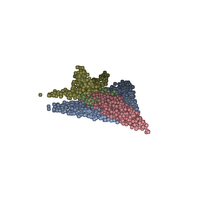}
\includegraphics[width=0.19\linewidth]{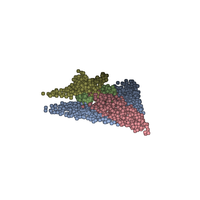}
\end{minipage}

\begin{minipage}{\linewidth}\centering
\includegraphics[width=0.19\linewidth]{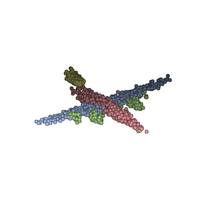}
\includegraphics[width=0.19\linewidth]{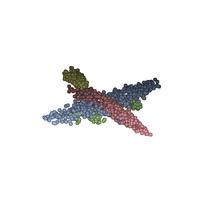}
\includegraphics[width=0.19\linewidth]{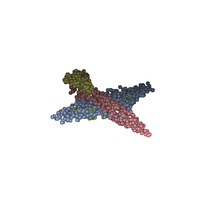}
\includegraphics[width=0.19\linewidth]{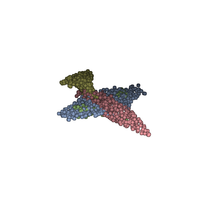}
\includegraphics[width=0.19\linewidth]{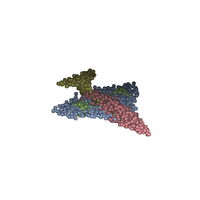}
\end{minipage}

\begin{minipage}{\linewidth}\centering
\includegraphics[width=0.19\linewidth]{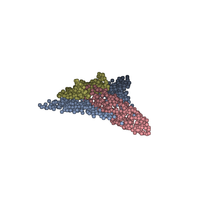}
\includegraphics[width=0.19\linewidth]{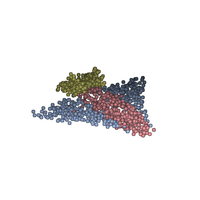}
\includegraphics[width=0.19\linewidth]{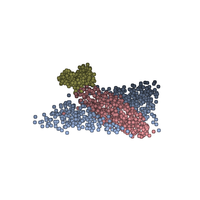}
\includegraphics[width=0.19\linewidth]{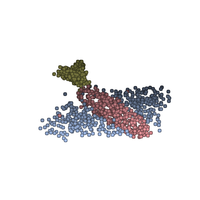}
\includegraphics[width=0.19\linewidth]{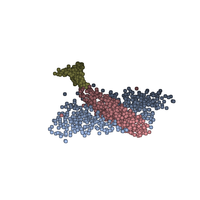}
\end{minipage}

\begin{minipage}{\linewidth}\centering
\includegraphics[width=0.19\linewidth]{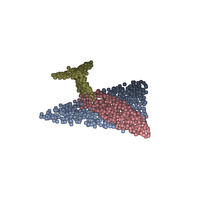}
\includegraphics[width=0.19\linewidth]{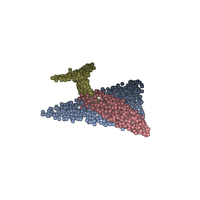}
\includegraphics[width=0.19\linewidth]{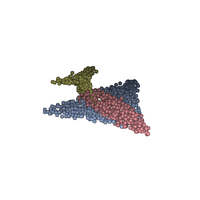}
\includegraphics[width=0.19\linewidth]{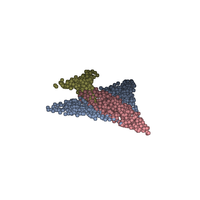}
\includegraphics[width=0.19\linewidth]{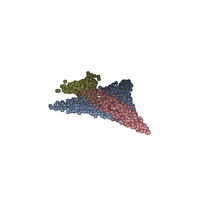}
\end{minipage}

\begin{minipage}{\linewidth}\centering
\includegraphics[width=0.19\linewidth]{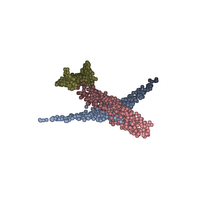}
\includegraphics[width=0.19\linewidth]{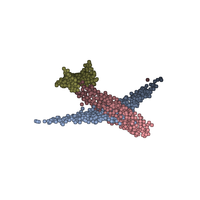}
\includegraphics[width=0.19\linewidth]{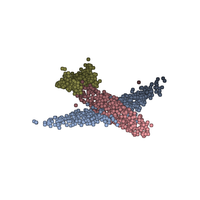}
\includegraphics[width=0.19\linewidth]{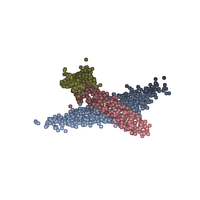}
\includegraphics[width=0.19\linewidth]{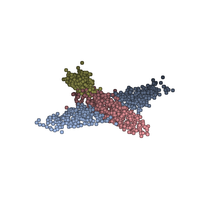}
\end{minipage}

	\caption{Complete latent space interpolation (parts and composition) for the airplane category. The shape's code on the left is linearly interpolated to the shape's code on the right.}
	\label{fig:supp_interpolation_plane_full}
	\end{figure*}

\begin{figure*}\centering
\begin{minipage}{\linewidth}\centering
\includegraphics[width=0.23\linewidth]{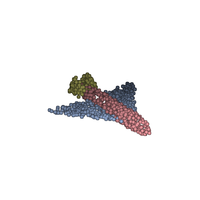}
\includegraphics[width=0.23\linewidth]{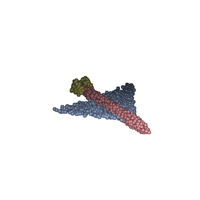}
\includegraphics[width=0.23\linewidth]{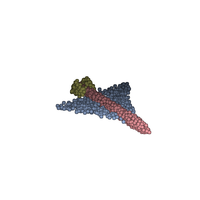}
\includegraphics[width=0.23\linewidth]{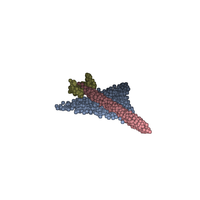}
\end{minipage}

\begin{minipage}{\linewidth}\centering
\includegraphics[width=0.23\linewidth]{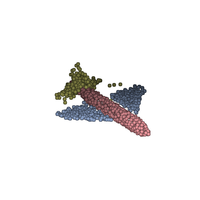}
\includegraphics[width=0.23\linewidth]{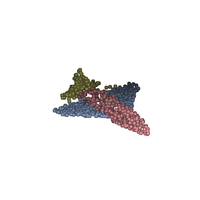}
\includegraphics[width=0.23\linewidth]{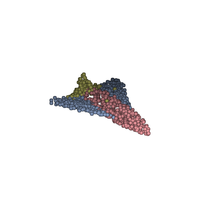}
\includegraphics[width=0.23\linewidth]{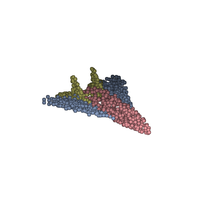}
\end{minipage}

\begin{minipage}{\linewidth}\centering
\includegraphics[width=0.23\linewidth]{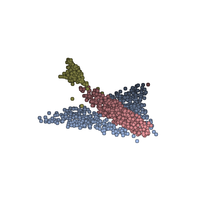}
\includegraphics[width=0.23\linewidth]{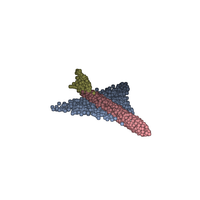}
\includegraphics[width=0.23\linewidth]{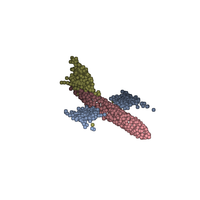}
\includegraphics[width=0.23\linewidth]{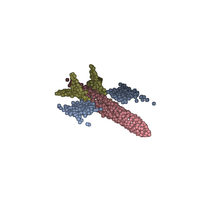}
\end{minipage}

\begin{minipage}{\linewidth}\centering
\includegraphics[width=0.23\linewidth]{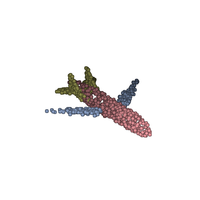}
\includegraphics[width=0.23\linewidth]{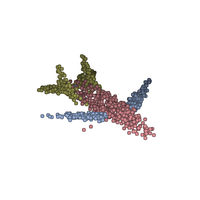}
\includegraphics[width=0.23\linewidth]{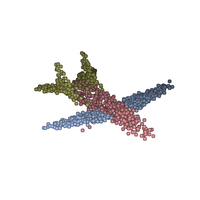}
\includegraphics[width=0.23\linewidth]{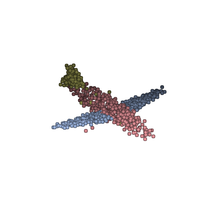}
\end{minipage}

	\caption{Airplanes part-by-part interpolation. In each row we present part-by-part interpolation between two airplanes (see left and right objects). The order of replacement is body-wings-tail.}
	\label{fig:supp_interpolation_plane_part}
	\end{figure*}

%% file: chairs_nn.tex
\begin{figure*}\centering
\begin{minipage}{0.48\linewidth}\centering
\begin{minipage}{\linewidth}\centering
\includegraphics[width=0.24\linewidth]{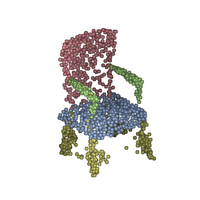}
\includegraphics[height=0.22\linewidth,width=0.02\linewidth]{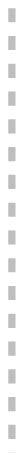}
\includegraphics[width=0.22\linewidth]{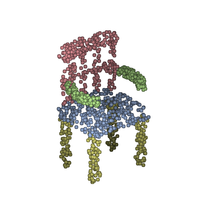}
\includegraphics[width=0.22\linewidth]{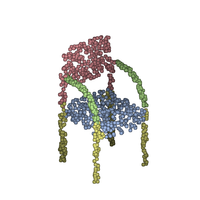}
\includegraphics[width=0.22\linewidth]{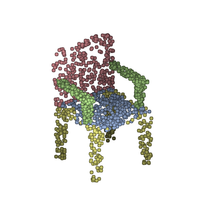}
\end{minipage}

\begin{minipage}{\linewidth}\centering
\includegraphics[width=0.24\linewidth]{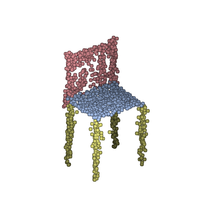}
\includegraphics[height=0.22\linewidth,width=0.02\linewidth]{figures/dashed_line.eps}
\includegraphics[width=0.22\linewidth]{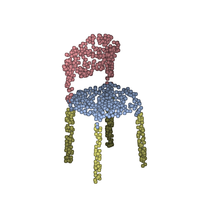}
\includegraphics[width=0.22\linewidth]{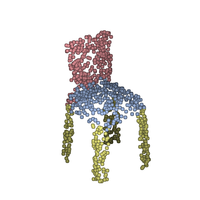}
\includegraphics[width=0.22\linewidth]{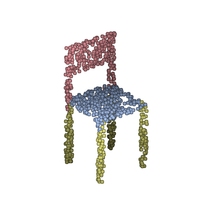}
\end{minipage}

\begin{minipage}{\linewidth}\centering
\includegraphics[width=0.24\linewidth]{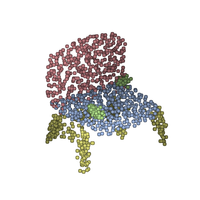}
\includegraphics[height=0.22\linewidth,width=0.02\linewidth]{figures/dashed_line.eps}
\includegraphics[width=0.22\linewidth]{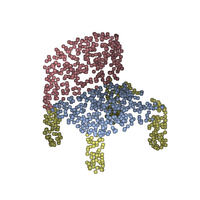}
\includegraphics[width=0.22\linewidth]{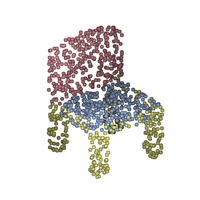}
\includegraphics[width=0.22\linewidth]{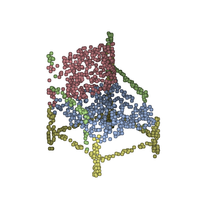}
\end{minipage}

\begin{minipage}{\linewidth}\centering
\includegraphics[width=0.24\linewidth]{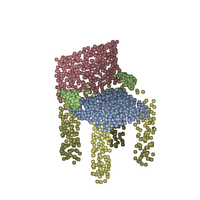}
\includegraphics[height=0.22\linewidth,width=0.02\linewidth]{figures/dashed_line.eps}
\includegraphics[width=0.22\linewidth]{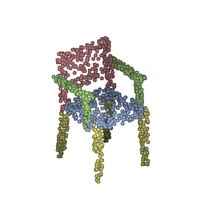}
\includegraphics[width=0.22\linewidth]{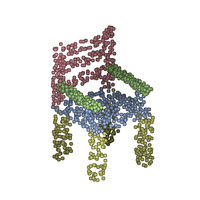}
\includegraphics[width=0.22\linewidth]{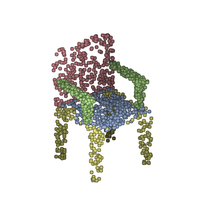}
\end{minipage}

\begin{minipage}{\linewidth}\centering
\includegraphics[width=0.24\linewidth]{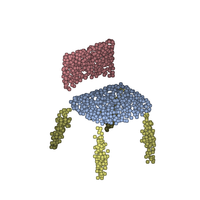}
\includegraphics[height=0.22\linewidth,width=0.02\linewidth]{figures/dashed_line.eps}
\includegraphics[width=0.22\linewidth]{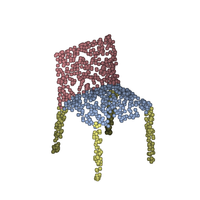}
\includegraphics[width=0.22\linewidth]{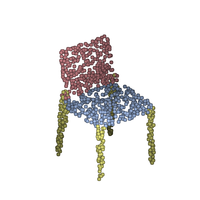}
\includegraphics[width=0.22\linewidth]{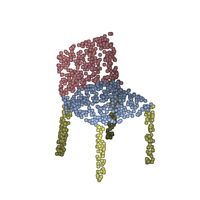}
\end{minipage}

\begin{minipage}{\linewidth}\centering
\includegraphics[width=0.24\linewidth]{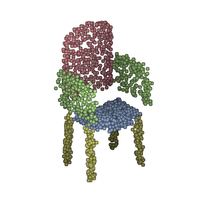}
\includegraphics[height=0.22\linewidth,width=0.02\linewidth]{figures/dashed_line.eps}
\includegraphics[width=0.22\linewidth]{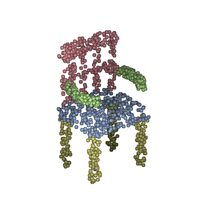}
\includegraphics[width=0.22\linewidth]{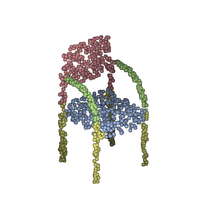}
\includegraphics[width=0.22\linewidth]{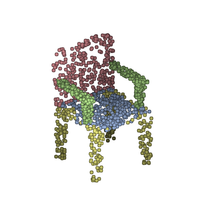}
\end{minipage}

\begin{minipage}{\linewidth}\centering
\includegraphics[width=0.24\linewidth]{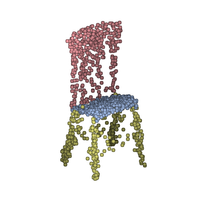}
\includegraphics[height=0.22\linewidth,width=0.02\linewidth]{figures/dashed_line.eps}
\includegraphics[width=0.22\linewidth]{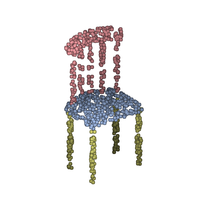}
\includegraphics[width=0.22\linewidth]{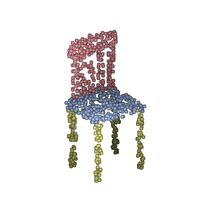}
\includegraphics[width=0.22\linewidth]{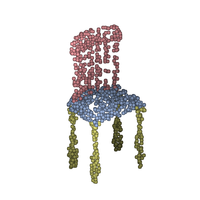}
\end{minipage}

\begin{minipage}{\linewidth}\centering
\includegraphics[width=0.24\linewidth]{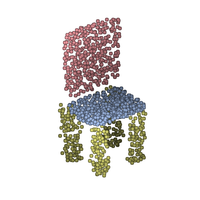}
\includegraphics[height=0.22\linewidth,width=0.02\linewidth]{figures/dashed_line.eps}
\includegraphics[width=0.22\linewidth]{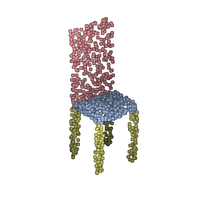}
\includegraphics[width=0.22\linewidth]{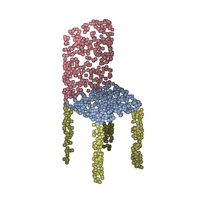}
\includegraphics[width=0.22\linewidth]{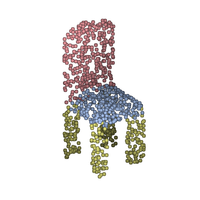}
\end{minipage}

\begin{minipage}{\linewidth}\centering
\includegraphics[width=0.24\linewidth]{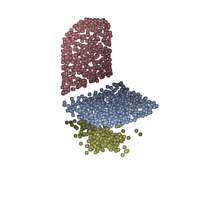}
\includegraphics[height=0.22\linewidth,width=0.02\linewidth]{figures/dashed_line.eps}
\includegraphics[width=0.22\linewidth]{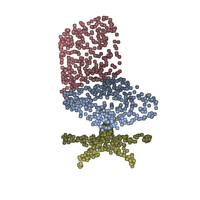}
\includegraphics[width=0.22\linewidth]{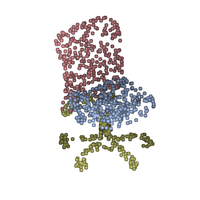}
\includegraphics[width=0.22\linewidth]{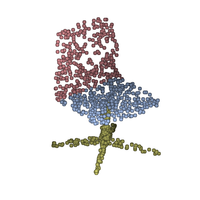}
\end{minipage}

\begin{minipage}{\linewidth}\centering
\includegraphics[width=0.24\linewidth]{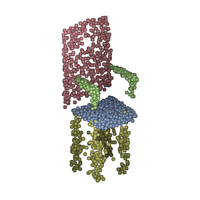}
\includegraphics[height=0.22\linewidth,width=0.02\linewidth]{figures/dashed_line.eps}
\includegraphics[width=0.22\linewidth]{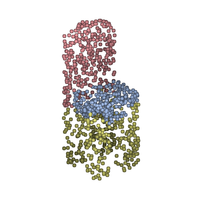}
\includegraphics[width=0.22\linewidth]{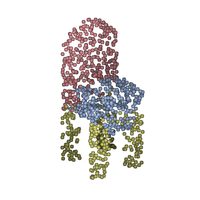}
\includegraphics[width=0.22\linewidth]{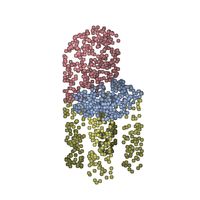}
\end{minipage}

\end{minipage}
\rulesep\begin{minipage}{0.48\linewidth}\centering
\begin{minipage}{\linewidth}\centering
\includegraphics[width=0.24\linewidth]{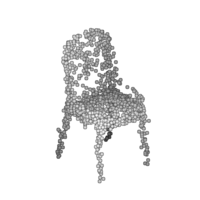}
\includegraphics[height=0.22\linewidth,width=0.02\linewidth]{figures/dashed_line.eps}
\includegraphics[width=0.22\linewidth]{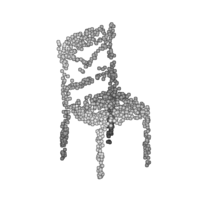}
\includegraphics[width=0.22\linewidth]{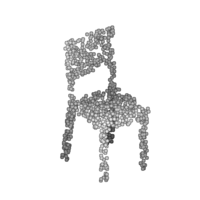}
\includegraphics[width=0.22\linewidth]{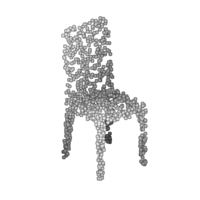}
\end{minipage}

\begin{minipage}{\linewidth}\centering
\includegraphics[width=0.24\linewidth]{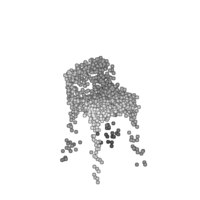}
\includegraphics[height=0.22\linewidth,width=0.02\linewidth]{figures/dashed_line.eps}
\includegraphics[width=0.22\linewidth]{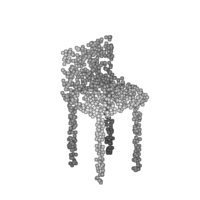}
\includegraphics[width=0.22\linewidth]{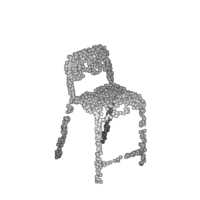}
\includegraphics[width=0.22\linewidth]{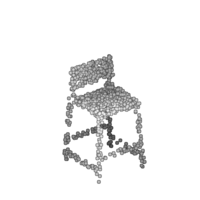}
\end{minipage}

\begin{minipage}{\linewidth}\centering
\includegraphics[width=0.24\linewidth]{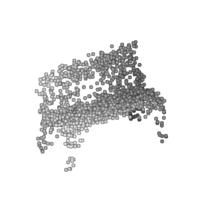}
\includegraphics[height=0.22\linewidth,width=0.02\linewidth]{figures/dashed_line.eps}
\includegraphics[width=0.22\linewidth]{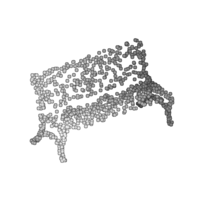}
\includegraphics[width=0.22\linewidth]{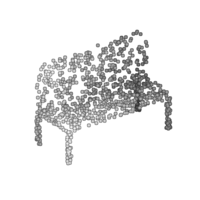}
\includegraphics[width=0.22\linewidth]{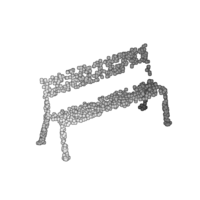}
\end{minipage}

\begin{minipage}{\linewidth}\centering
\includegraphics[width=0.24\linewidth]{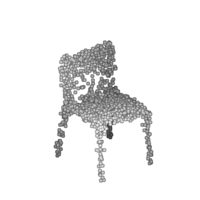}
\includegraphics[height=0.22\linewidth,width=0.02\linewidth]{figures/dashed_line.eps}
\includegraphics[width=0.22\linewidth]{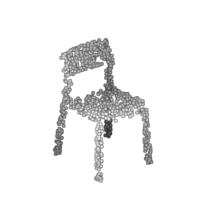}
\includegraphics[width=0.22\linewidth]{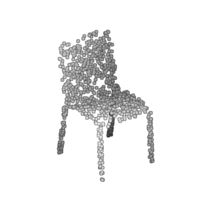}
\includegraphics[width=0.22\linewidth]{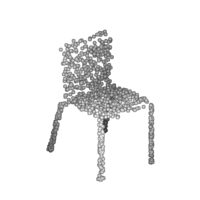}
\end{minipage}

\begin{minipage}{\linewidth}\centering
\includegraphics[width=0.24\linewidth]{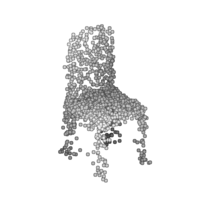}
\includegraphics[height=0.22\linewidth,width=0.02\linewidth]{figures/dashed_line.eps}
\includegraphics[width=0.22\linewidth]{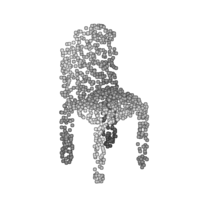}
\includegraphics[width=0.22\linewidth]{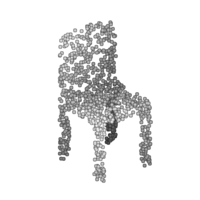}
\includegraphics[width=0.22\linewidth]{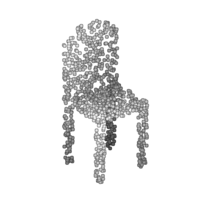}
\end{minipage}

\begin{minipage}{\linewidth}\centering
\includegraphics[width=0.24\linewidth]{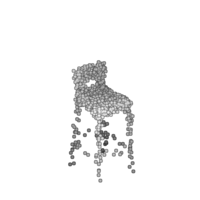}
\includegraphics[height=0.22\linewidth,width=0.02\linewidth]{figures/dashed_line.eps}
\includegraphics[width=0.22\linewidth]{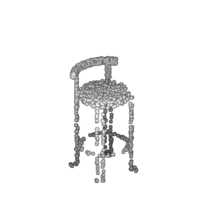}
\includegraphics[width=0.22\linewidth]{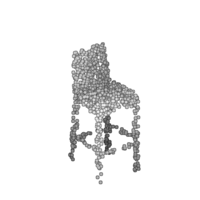}
\includegraphics[width=0.22\linewidth]{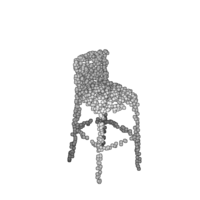}
\end{minipage}

\begin{minipage}{\linewidth}\centering
\includegraphics[width=0.24\linewidth]{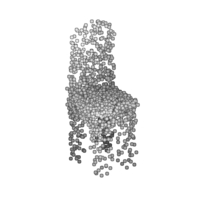}
\includegraphics[height=0.22\linewidth,width=0.02\linewidth]{figures/dashed_line.eps}
\includegraphics[width=0.22\linewidth]{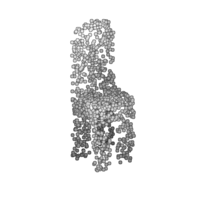}
\includegraphics[width=0.22\linewidth]{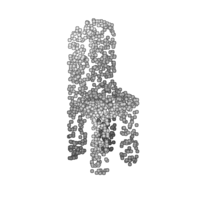}
\includegraphics[width=0.22\linewidth]{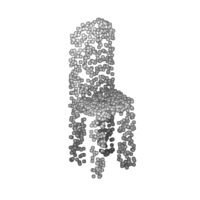}
\end{minipage}

\begin{minipage}{\linewidth}\centering
\includegraphics[width=0.24\linewidth]{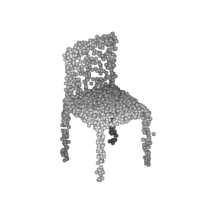}
\includegraphics[height=0.22\linewidth,width=0.02\linewidth]{figures/dashed_line.eps}
\includegraphics[width=0.22\linewidth]{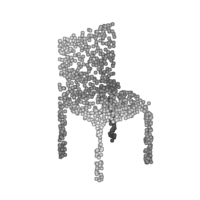}
\includegraphics[width=0.22\linewidth]{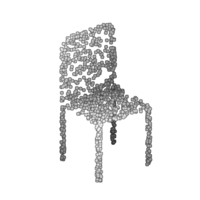}
\includegraphics[width=0.22\linewidth]{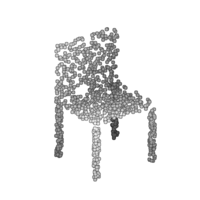}
\end{minipage}

\begin{minipage}{\linewidth}\centering
\includegraphics[width=0.24\linewidth]{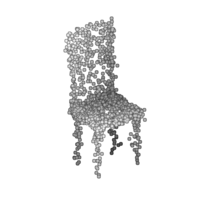}
\includegraphics[height=0.22\linewidth,width=0.02\linewidth]{figures/dashed_line.eps}
\includegraphics[width=0.22\linewidth]{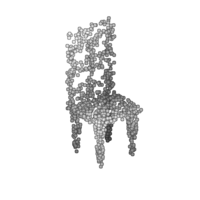}
\includegraphics[width=0.22\linewidth]{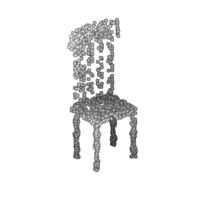}
\includegraphics[width=0.22\linewidth]{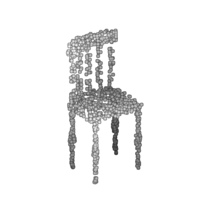}
\end{minipage}

\begin{minipage}{\linewidth}\centering
\includegraphics[width=0.24\linewidth]{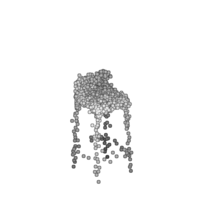}
\includegraphics[height=0.22\linewidth,width=0.02\linewidth]{figures/dashed_line.eps}
\includegraphics[width=0.22\linewidth]{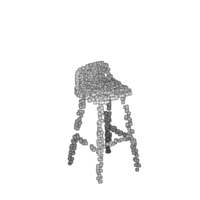}
\includegraphics[width=0.22\linewidth]{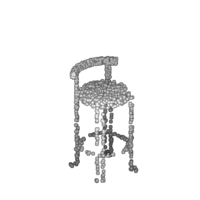}
\includegraphics[width=0.22\linewidth]{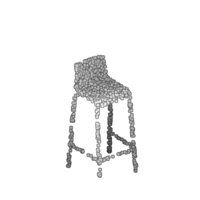}
\end{minipage}

\end{minipage}
	\caption{Chairs nearest-neighbors. We present random generated results from \OurNet{} and the baseline, for each result, we present its three nearest-neighbors based on the Chamfer distance. In the left half we show results from \OurNet{} (in colors), where the left column is generated and the next three columns to the right are its three nearest-neighbors. In the right half we present results from the baseline (gray), where the left column is generated and the next three columns to the right are its three nearest-neighbors.}
	\label{fig:chairs_nn}
	\end{figure*}

%% file: plane_nn.tex
\begin{figure*}\centering
\begin{minipage}{0.48\linewidth}\centering
\begin{minipage}{\linewidth}\centering
\includegraphics[width=0.24\linewidth]{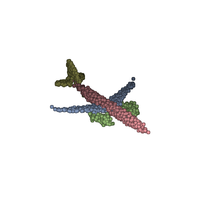}
\includegraphics[height=0.22\linewidth,width=0.02\linewidth]{figures/dashed_line.eps}
\includegraphics[width=0.22\linewidth]{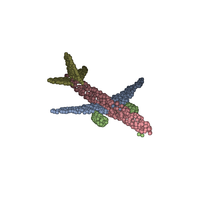}
\includegraphics[width=0.22\linewidth]{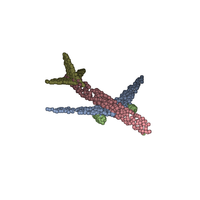}
\includegraphics[width=0.22\linewidth]{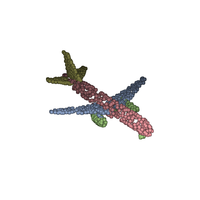}
\end{minipage}

\begin{minipage}{\linewidth}\centering
\includegraphics[width=0.24\linewidth]{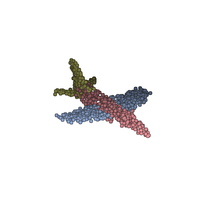}
\includegraphics[height=0.22\linewidth,width=0.02\linewidth]{figures/dashed_line.eps}
\includegraphics[width=0.22\linewidth]{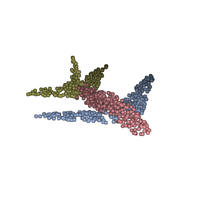}
\includegraphics[width=0.22\linewidth]{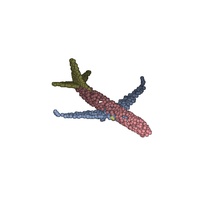}
\includegraphics[width=0.22\linewidth]{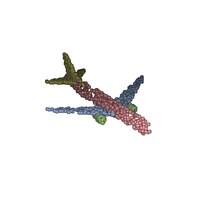}
\end{minipage}

\begin{minipage}{\linewidth}\centering
\includegraphics[width=0.24\linewidth]{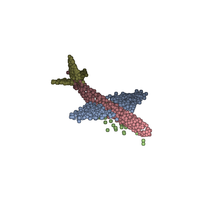}
\includegraphics[height=0.22\linewidth,width=0.02\linewidth]{figures/dashed_line.eps}
\includegraphics[width=0.22\linewidth]{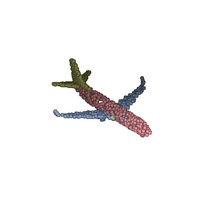}
\includegraphics[width=0.22\linewidth]{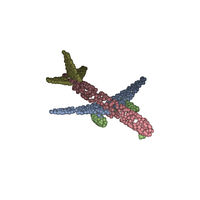}
\includegraphics[width=0.22\linewidth]{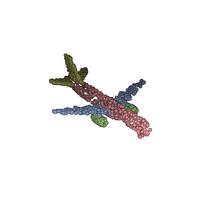}
\end{minipage}

\begin{minipage}{\linewidth}\centering
\includegraphics[width=0.24\linewidth]{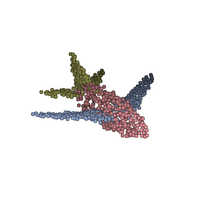}
\includegraphics[height=0.22\linewidth,width=0.02\linewidth]{figures/dashed_line.eps}
\includegraphics[width=0.22\linewidth]{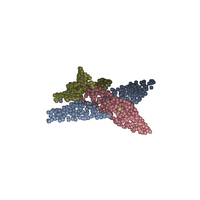}
\includegraphics[width=0.22\linewidth]{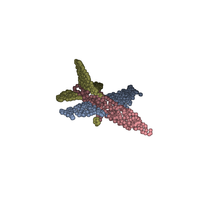}
\includegraphics[width=0.22\linewidth]{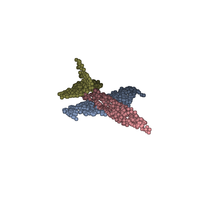}
\end{minipage}

\begin{minipage}{\linewidth}\centering
\includegraphics[width=0.24\linewidth]{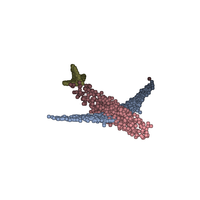}
\includegraphics[height=0.22\linewidth,width=0.02\linewidth]{figures/dashed_line.eps}
\includegraphics[width=0.22\linewidth]{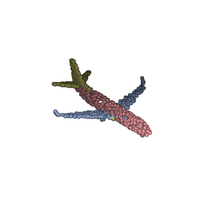}
\includegraphics[width=0.22\linewidth]{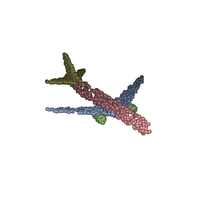}
\includegraphics[width=0.22\linewidth]{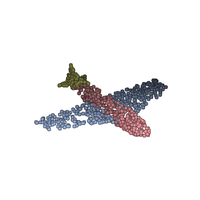}
\end{minipage}

\begin{minipage}{\linewidth}\centering
\includegraphics[width=0.24\linewidth]{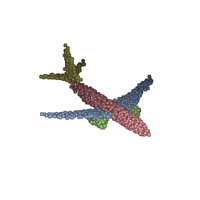}
\includegraphics[height=0.22\linewidth,width=0.02\linewidth]{figures/dashed_line.eps}
\includegraphics[width=0.22\linewidth]{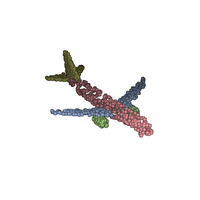}
\includegraphics[width=0.22\linewidth]{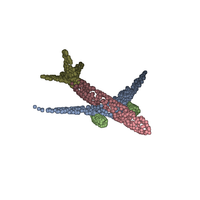}
\includegraphics[width=0.22\linewidth]{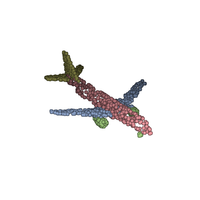}
\end{minipage}

\begin{minipage}{\linewidth}\centering
\includegraphics[width=0.24\linewidth]{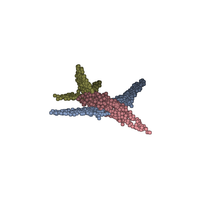}
\includegraphics[height=0.22\linewidth,width=0.02\linewidth]{figures/dashed_line.eps}
\includegraphics[width=0.22\linewidth]{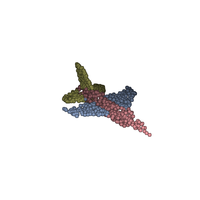}
\includegraphics[width=0.22\linewidth]{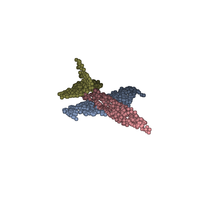}
\includegraphics[width=0.22\linewidth]{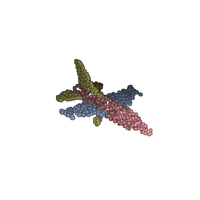}
\end{minipage}

\begin{minipage}{\linewidth}\centering
\includegraphics[width=0.24\linewidth]{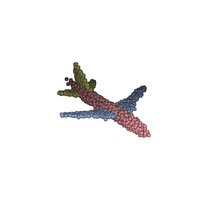}
\includegraphics[height=0.22\linewidth,width=0.02\linewidth]{figures/dashed_line.eps}
\includegraphics[width=0.22\linewidth]{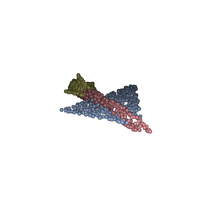}
\includegraphics[width=0.22\linewidth]{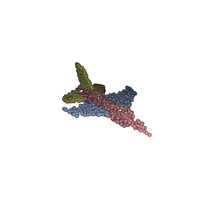}
\includegraphics[width=0.22\linewidth]{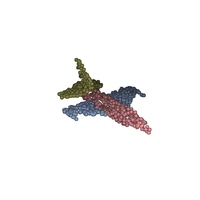}
\end{minipage}

\begin{minipage}{\linewidth}\centering
\includegraphics[width=0.24\linewidth]{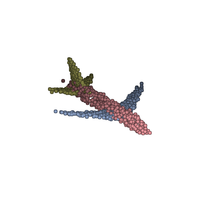}
\includegraphics[height=0.22\linewidth,width=0.02\linewidth]{figures/dashed_line.eps}
\includegraphics[width=0.22\linewidth]{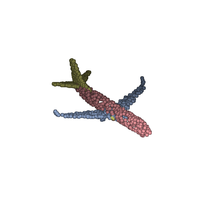}
\includegraphics[width=0.22\linewidth]{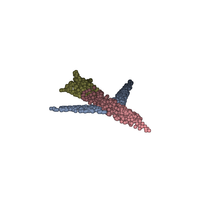}
\includegraphics[width=0.22\linewidth]{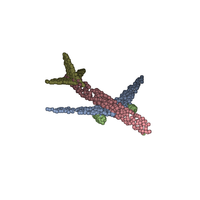}
\end{minipage}

\begin{minipage}{\linewidth}\centering
\includegraphics[width=0.24\linewidth]{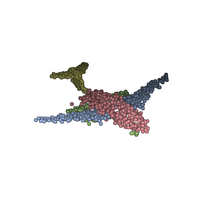}
\includegraphics[height=0.22\linewidth,width=0.02\linewidth]{figures/dashed_line.eps}
\includegraphics[width=0.22\linewidth]{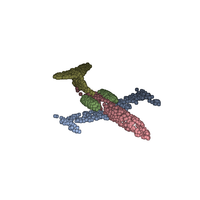}
\includegraphics[width=0.22\linewidth]{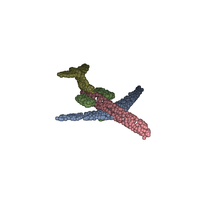}
\includegraphics[width=0.22\linewidth]{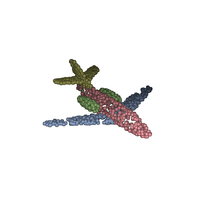}
\end{minipage}

\end{minipage}
\rulesep\begin{minipage}{0.48\linewidth}\centering
\begin{minipage}{\linewidth}\centering
\includegraphics[width=0.24\linewidth]{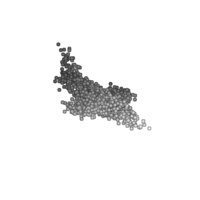}
\includegraphics[height=0.22\linewidth,width=0.02\linewidth]{figures/dashed_line.eps}
\includegraphics[width=0.22\linewidth]{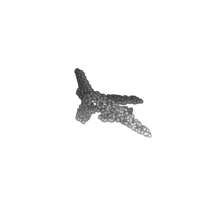}
\includegraphics[width=0.22\linewidth]{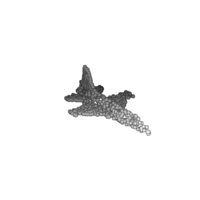}
\includegraphics[width=0.22\linewidth]{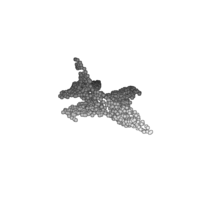}
\end{minipage}

\begin{minipage}{\linewidth}\centering
\includegraphics[width=0.24\linewidth]{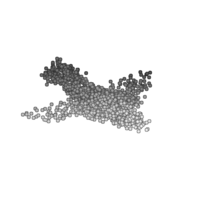}
\includegraphics[height=0.22\linewidth,width=0.02\linewidth]{figures/dashed_line.eps}
\includegraphics[width=0.22\linewidth]{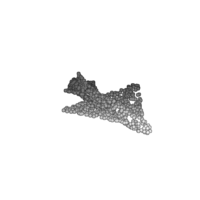}
\includegraphics[width=0.22\linewidth]{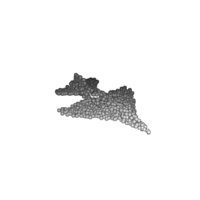}
\includegraphics[width=0.22\linewidth]{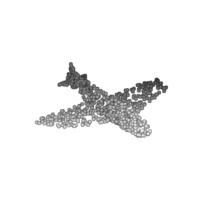}
\end{minipage}

\begin{minipage}{\linewidth}\centering
\includegraphics[width=0.24\linewidth]{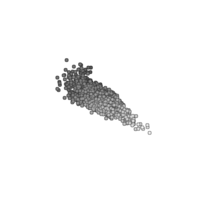}
\includegraphics[height=0.22\linewidth,width=0.02\linewidth]{figures/dashed_line.eps}
\includegraphics[width=0.22\linewidth]{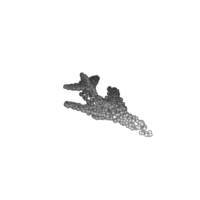}
\includegraphics[width=0.22\linewidth]{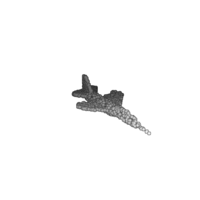}
\includegraphics[width=0.22\linewidth]{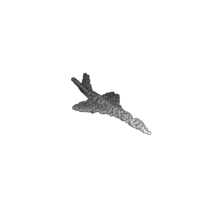}
\end{minipage}

\begin{minipage}{\linewidth}\centering
\includegraphics[width=0.24\linewidth]{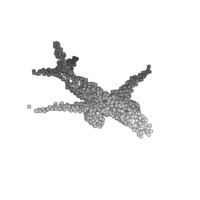}
\includegraphics[height=0.22\linewidth,width=0.02\linewidth]{figures/dashed_line.eps}
\includegraphics[width=0.22\linewidth]{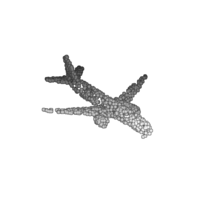}
\includegraphics[width=0.22\linewidth]{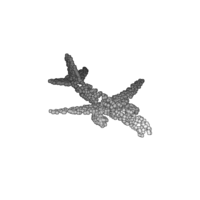}
\includegraphics[width=0.22\linewidth]{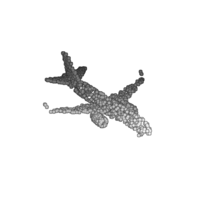}
\end{minipage}

\begin{minipage}{\linewidth}\centering
\includegraphics[width=0.24\linewidth]{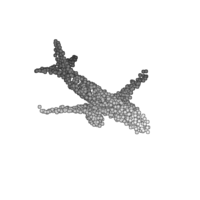}
\includegraphics[height=0.22\linewidth,width=0.02\linewidth]{figures/dashed_line.eps}
\includegraphics[width=0.22\linewidth]{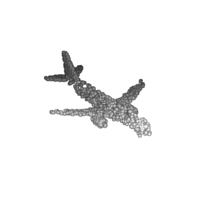}
\includegraphics[width=0.22\linewidth]{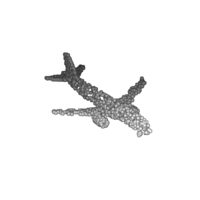}
\includegraphics[width=0.22\linewidth]{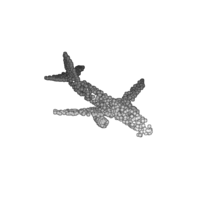}
\end{minipage}

\begin{minipage}{\linewidth}\centering
\includegraphics[width=0.24\linewidth]{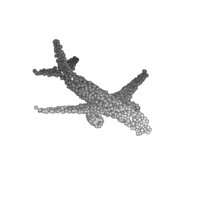}
\includegraphics[height=0.22\linewidth,width=0.02\linewidth]{figures/dashed_line.eps}
\includegraphics[width=0.22\linewidth]{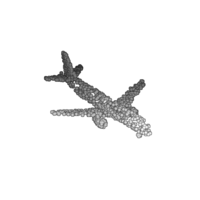}
\includegraphics[width=0.22\linewidth]{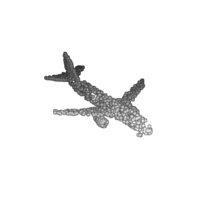}
\includegraphics[width=0.22\linewidth]{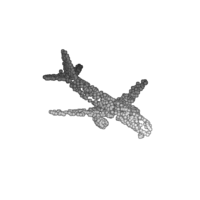}
\end{minipage}

\begin{minipage}{\linewidth}\centering
\includegraphics[width=0.24\linewidth]{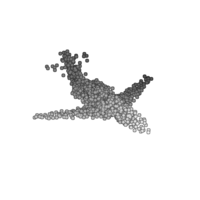}
\includegraphics[height=0.22\linewidth,width=0.02\linewidth]{figures/dashed_line.eps}
\includegraphics[width=0.22\linewidth]{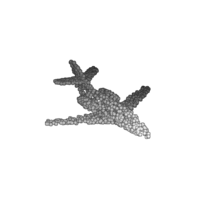}
\includegraphics[width=0.22\linewidth]{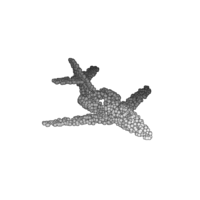}
\includegraphics[width=0.22\linewidth]{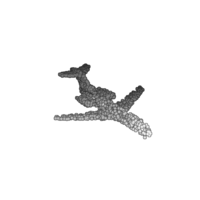}
\end{minipage}

\begin{minipage}{\linewidth}\centering
\includegraphics[width=0.24\linewidth]{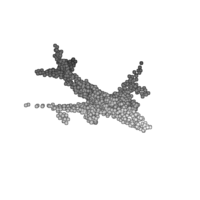}
\includegraphics[height=0.22\linewidth,width=0.02\linewidth]{figures/dashed_line.eps}
\includegraphics[width=0.22\linewidth]{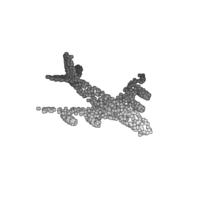}
\includegraphics[width=0.22\linewidth]{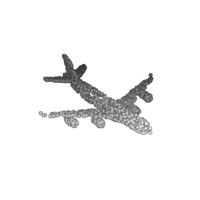}
\includegraphics[width=0.22\linewidth]{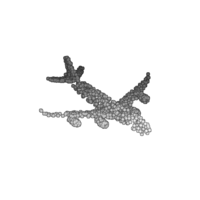}
\end{minipage}

\begin{minipage}{\linewidth}\centering
\includegraphics[width=0.24\linewidth]{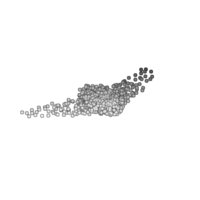}
\includegraphics[height=0.22\linewidth,width=0.02\linewidth]{figures/dashed_line.eps}
\includegraphics[width=0.22\linewidth]{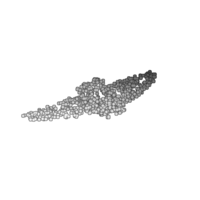}
\includegraphics[width=0.22\linewidth]{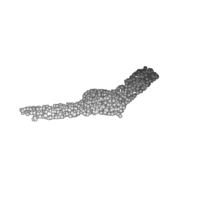}
\includegraphics[width=0.22\linewidth]{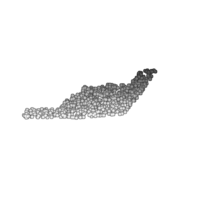}
\end{minipage}

\begin{minipage}{\linewidth}\centering
\includegraphics[width=0.24\linewidth]{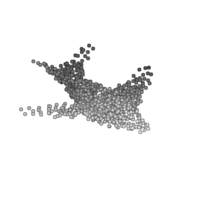}
\includegraphics[height=0.22\linewidth,width=0.02\linewidth]{figures/dashed_line.eps}
\includegraphics[width=0.22\linewidth]{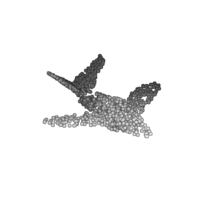}
\includegraphics[width=0.22\linewidth]{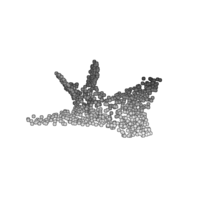}
\includegraphics[width=0.22\linewidth]{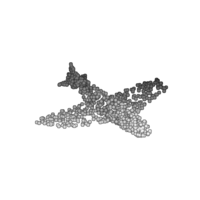}
\end{minipage}

\end{minipage}
	\caption{Airplanes nearest-neighbors. We present random generated results from \OurNet{} and the baseline, for each result, we present its three nearest-neighbors based on the Chamfer distance. In the left half we show results from \OurNet{} (in colors), where the left column is generated and the next three columns to the right are its three nearest-neighbors. In the right half we present results from the baseline (gray), where the left column is generated and the next three columns to the right are its three nearest-neighbors.}
	\label{fig:plane_nn}
	\end{figure*}